\documentclass[]{aastex61}

\usepackage{amsmath,amsfonts,isomath}
\usepackage[american]{babel}

\usepackage{soul}
\usepackage{xcolor}
\newcommand{\revadd}[1] {#1}
\newcommand{\revrem}[1] {}

\usepackage{hyperref}

\begin{document}

\title{A Compressed Sensing-based Image Reconstruction Algorithm for Solar Flare X-Ray Observations}
\shorttitle{Compressed Sensing-based Image Reconstruction}

\author[0000-0002-3979-128X]{Simon Felix}
\affiliation{University of Applied Sciences and Arts Northwestern Switzerland FHNW, 5210 Windisch, Switzerland}
\email{simon.felix@fhnw.ch}
\author[0000-0001-8633-8819]{Roman Bolzern}
\affiliation{University of Applied Sciences and Arts Northwestern Switzerland FHNW, 5210 Windisch, Switzerland}
\email{roman.bolzern@fhnw.ch}
\author[0000-0003-1438-9099]{Marina Battaglia}
\affiliation{University of Applied Sciences and Arts Northwestern Switzerland FHNW, 5210 Windisch, Switzerland}
\email{marina.battaglia@fhnw.ch}

\begin{abstract}
One way of imaging X-ray emission from solar flares is to measure Fourier components of the spatial X-ray source distribution. We present a new Compressed Sensing-based algorithm named VIS\_CS, which reconstructs the spatial distribution from such Fourier components. We demonstrate the application of the algorithm on synthetic and observed solar flare X-ray data from the Reuven Ramaty High Energy Solar Spectroscopic Imager (\emph{RHESSI}) satellite and compare its performance with existing algorithms. VIS\_CS produces competitive results with accurate photometry and morphology, without requiring any algorithm- and X-ray source-specific parameter tuning. Its robustness and performance make this algorithm ideally suited for generation of quicklook images or large image cubes without user intervention, such as for imaging spectroscopy analysis.
\end{abstract}


\section{Introduction}
Solar flares are the most energetic phenomena in the solar system, \revadd{releasing energy of up to 10$^{25}$ J in seconds to minutes converting it to accelerated particles and plasma heating and generating emission across the electromagnetic spectrum in the process.} X-rays and gamma-rays are the most direct signature of flare-accelerated particles and provide valuable insights which are central for the understanding of the underlying physical processes (see \citet{Fle11} for a review). Due to their high energy, X-ray photons cannot be focused easily with traditional optics. A widely used method in solar physics to image high energy X-rays is indirect Fourier imaging. The X-ray flux is modulated temporally and/or spatially to measure the complex Fourier components of the source distribution. The resulting data products (so-called \emph{visibilities}) are mathematically equivalent to the data products obtained by radio interferometers \citep{2002SoPh..210...61H}.

The most recent X-ray imager employing this method is the NASA spacecraft Reuven Ramaty High Energy Solar Spectroscopic Imager (\emph{RHESSI}, \citet{lin2003reuven}). This instrument consists of a telescope holding nine Germanium detectors, each behind \revadd{a} bi-grid, \revadd{a so-called ``subcollimator''}, with different pitch.
As the spacecraft spins around its own axis, \revadd{the subcollimators modulate the X-ray flux temporally, which is measured by the Germanium detector behind each subcollimator.} Several algorithms have been developed in the past to reconstruct an image from these measurements. The individual strengths and weaknesses of existing algorithms are discussed in section \ref{sec:existing}. All of these algorithms typically require algorithm- and source-specific parameter tuning and detector selection to produce best results, something that requires in-depth knowledge of the instrument and image algorithms.

\begin{figure*}[t]
\includegraphics[width=0.99\textwidth]{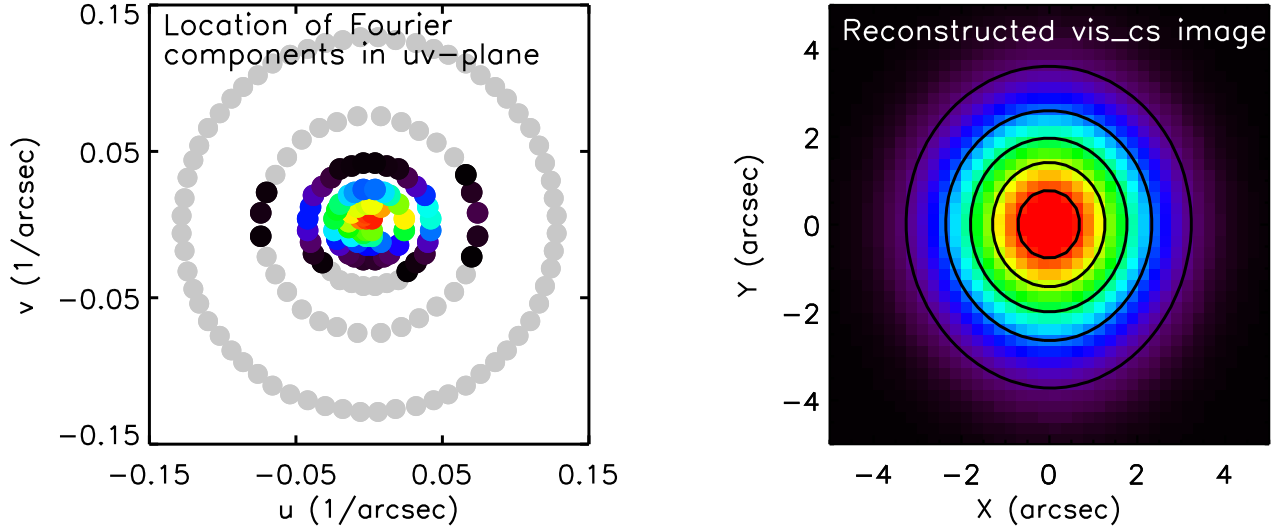}
\caption{Illustration of image reconstruction from complex Fourier components \revadd{as measured by the \emph{RHESSI} satellite}. Left: Locations of Fourier components (visibilities) of a simple two-dimensional Gaussian source in the $uv$-plane, calculated from the time modulated X-ray flux. Each circle corresponds to one of 9 \emph{RHESSI} \revrem{bi-grids}\revadd{subcollimators}. Grey dots indicate visibilities with a weak signal. Right: Reconstructed image using the VIS\_CS algorithm. The black isolines mark 10\%, 30\%, 50\%, 70\%, and 90\% of peak emission.}
\label{fig:vi_ill}
\end{figure*}

One particular challenge of Fourier imaging is that the number of measured visibilities is necessarily limited by photon statistics and fabrication limits of the \revadd{subcollimators}. Images are reconstructed typically from 50 to 300 visibilities (Figure~\ref{fig:vi_ill}). The Nyquist-Shannon sampling theorem states that an arbitrary signal with bandwidth $b$ must be sampled at $\geqslant 2b$ points to allow the reconstruction of the original signal \citep{landau1967necessary}. For the problem at hand this means that at least $2b$ visibilities are required to reconstruct an image with $2b$ pixels perfectly. As \emph{RHESSI} records far fewer visibilities than what would be required for even the smallest of reconstructions (e.g. $64 \times 64 = 4096$ pixels from less than 300 visibilities), perfect reconstructions are fundamentally impossible. High-resolution image reconstruction from only a few measured visibilities is an under-de\-ter\-mined, ill-posed problem. There are infinitely many reconstructions that fit the measured visibilities, thus any specific reconstruction is inherently biased in some way. All sparse image reconstruction algorithms face this problem.

Here we present a new algorithm (VIS\_CS) that is based on the fact that X-ray source distributions are \emph{not} arbitrary signals, but sparse when represented in an appropriate basis. Most natural signals are sparse in some bases. \citet{Can06} \revadd{and \citet{donoho2006compressed}} showed first that sparse signals can be reconstructed from far fewer samples than the Nyquist-Shannon sampling theorem suggests. \revrem{This is called "Compressed Sensing".}\revadd{This so-called ``Compressed Sensing'' is possible when the signal, the measurement device and a suitable reconstruction algorithm all satisfy several constraints, as described in the previously mentioned papers. Despite the fact that \emph{RHESSI} violates these constraints, Compressed Sensing reconstruction algorithms} \revrem{work}\revadd{can be applied in this setting. Given that the \emph{RHESSI} instrument cannot be changed, we solely concentrate on the reconstruction algorithm.}

In this paper we show the application of Compressed Sensing to the image reconstruction problem \revrem{of}\revadd{from} \emph{RHESSI} visibilities. We compare and contrast our approach with other image reconstruction algorithms and show that less parameter tuning is required to obtain plausible reconstructions, making the algorithm particularly useful for automatic generation of quick-look images and large image-cubes for imaging spectroscopy. In Section \ref{sec:existing} we give a brief overview of existing imaging algorithms for \emph{RHESSI} observations. The VIS\_CS algorithm is described in Section \ref{sec:alg}. Tests of the algorithm using synthetic data, as well as solar flare data observed with \emph{RHESSI}, are presented in Section \ref{sec:test}. 


\section{Existing image reconstruction methods}\label{sec:existing}
Several reconstruction methods for X-ray images were first developed for visibilities recorded by radio interferometry \revadd{\citep{thompson2001interferometry}}. Before the \emph{RHESSI} mission, such reconstruction methods were applied to the hard X-ray Telescope HXT of the Japanese Yohkoh spacecraft \citet{kosugi1991hard}. Those reconstruction methods have been adopted by the \emph{RHESSI} mission and have since been constantly improved upon. Some algorithms have been developed specifically for \emph{RHESSI}. Here we give a brief overview of the most commonly used algorithms.

\paragraph{Back projection} The back projection algorithm \citep{2002SoPh..210...61H} performs an inverse Fourier transform where all non-measured Fourier components are assumed to be zero. The resulting reconstructions show heavy ringing artefacts and contain areas with unphysical negative flux. \revrem{This algorithm is the simplest reconstruction algorithm.}\revadd{Note that the standard \emph{RHESSI} implementation of this simple algorithm is based on time-modulated photon counts instead of visibilities.}

\paragraph{Clean} This algorithm iteratively subtracts point sources, convolved with the point spread function of the instrument, from the backprojected image. This process is repeated until the residuals are below a predefined threshold \citep{Hog74,2002SoPh..210...61H}. Clean \revrem{always yields an image, and} rarely produces spurious or over-resolved sources. However, to resolve small sources users have to carefully select the \revrem{grids}\revadd{subcollimators} that are used for the reconstruction (and thus the spatial resolution) and Clean beam width. It is currently the standard algorithm for reconstructions of large numbers of images and generation of quicklook images.

\paragraph{uv\_smooth} The uv\_smooth algorithm reconstructs missing Fourier components by interpolating the visibilities in the $uv$-plane \citep{alM09}. It then iteratively backprojects visibilities and thresholds the image to impose a positivity constraint. uv\_smooth is the fastest algorithm currently in use, but produces ringing artifacts in some cases, as is apparent in the aforementioned publication by the authors of uv\_smooth.

\paragraph{Pixon} \citet{Met96} assume that images are superpositions of circular sources (\emph{pixons}). Pixon employs a specialized \revrem{search procedure}\revadd{heuristic} to balance the degrees of freedom (the number of pixons) while maintaining consistency with the data. The Pixon algorithm generally provides excellent photometry, but can be computationally demanding and suffers from over-resolution artifacts when not tuned, as reported by \citet{Kru11}. \revadd{Pixon shares some ideas with our algorithm as it exploits sparsity for its reconstructions as well. But Pixon uses simpler basis functions, a different approach to solve the optimization problem and it operates on time-modulated photon counts instead of visibilities.}

\paragraph{Forward Fit} This algorithm performs, as its name indicates, a parametric fit of a model to the measured data \citep{aschwanden2002reconstruction}. A priori, users choose a model to be fitted to the source distribution. The algorithm supports models with up to three two-dimensional, circular or elliptic, Gaussians. If the chosen model matches reality reasonably well, Forward fit produces good results. Two versions exist: The original implementation operates on photon counts, a more recent implementation operates on visibilities (VIS\_FWDFIT). This newer implementation is widely used to determine the centers of mass of sources as it is capable of measuring these with sub-arcsecond accuracy and provides uncertainties.

\section{Image reconstruction with Compressed Sensing} \label{sec:alg}
Our algorithm VIS\_CS generates plausible reconstructions, consistent with the measurements, by superimposing Gaussian distributions. Let $ \mathbfit{x} \in \mathbb{R}^n$ be the spatial source distribution in image space we wish to reconstruct. Without loss of generality, we assume that the source distribution can be represented discretely as vector with $ n $ elements (i.e. pixels). $\mathbfit{y} \in \mathbb{C}^m$ represents the $m$ measured visibilities and $\mathbfit{\eta} \in \mathbb{C}^m$ is a noise vector. The matrix $\mathbf{A} \in \mathbb{C}^{m \times n}$ represents the linear transformation that is performed physically by \emph{RHESSI}'s modulating \revrem{grids}\revadd{subcollimators} and detectors that transform the spatial flux distribution into visibilities. \emph{RHESSI} observes $\mathbfit{x}$ as noisy visibilities $\mathbfit{y} = \mathbf{A}\mathbfit{x} + \mathbfit{\eta}$. Given $\mathbf{A}$ and $\mathbfit{y}$, an image reconstruction method tries to find a reconstruction $\mathbfit{\hat{x}}$ that is close to $\mathbfit{x}$:
\begin{equation}
\begin{aligned}
& \underset{\mathbfit{\hat{x}}}{\text{minimize}}
& & \left\lVert \mathbf{A}\mathbfit{\hat{x}}-\mathbfit{y} \right\rVert \\
\end{aligned}
\end{equation}
Compressed Sensing \citep{Can06} shows that signals $\mathbfit{x}$ can be reconstructed from relatively few measurements ($m \ll n$), if the signal is sufficiently sparse when represented in an appropriate basis. Suitable sparse bases may be overcomplete, i.e. the basis may contain many more basis functions than strictly necessary. Also, the basis functions need not be orthogonal. We use matrix $\mathbf{D} \in \mathbb{R}^{m \times n}$ to describe the linear transformation from a suitable sparse basis to spatial image space. If the Compressed Sensing assumption holds and $\mathbfit{x}$ is sparse in the chosen basis, we may add $\mathbf{D}\mathbfit{\hat{x}}$ as $\ell_{0}$ regularizer to bias the solution towards sparse reconstructions:

\begin{equation}
\begin{aligned}
& \underset{\hat{x}}{\text{minimize}}
& & (1-\lambda)\left\lVert \mathbf{A}\mathbfit{\hat{x}}-\mathbfit{y} \right\rVert_2^2 + \lambda\|\mathbf{D}\mathbfit{\hat{x}}\|_0 \\
\end{aligned}
\end{equation}

\begin{figure*}
\begin{tabular}{ccccc}
\includegraphics[width=0.176\textwidth]{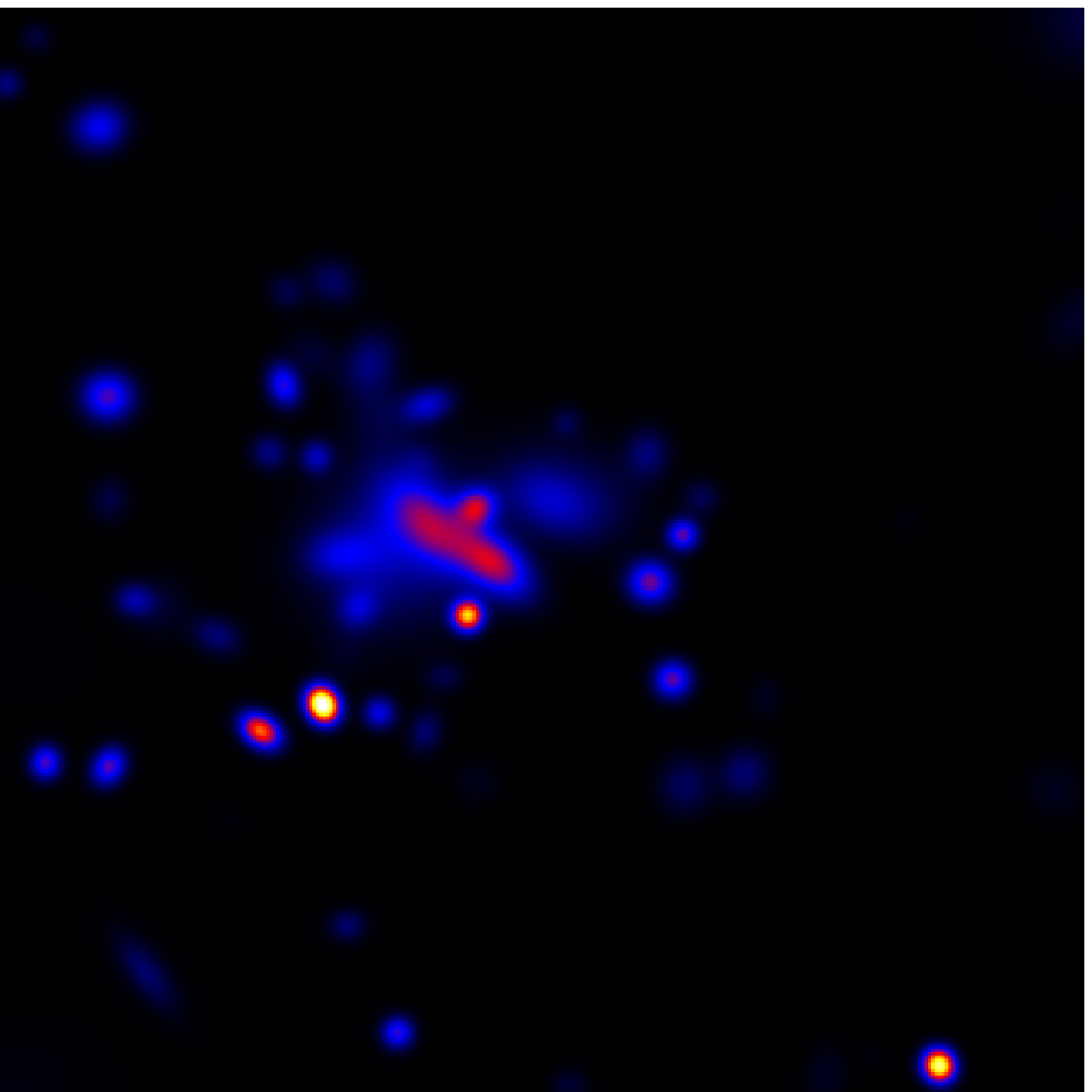} &
\includegraphics[width=0.176\textwidth]{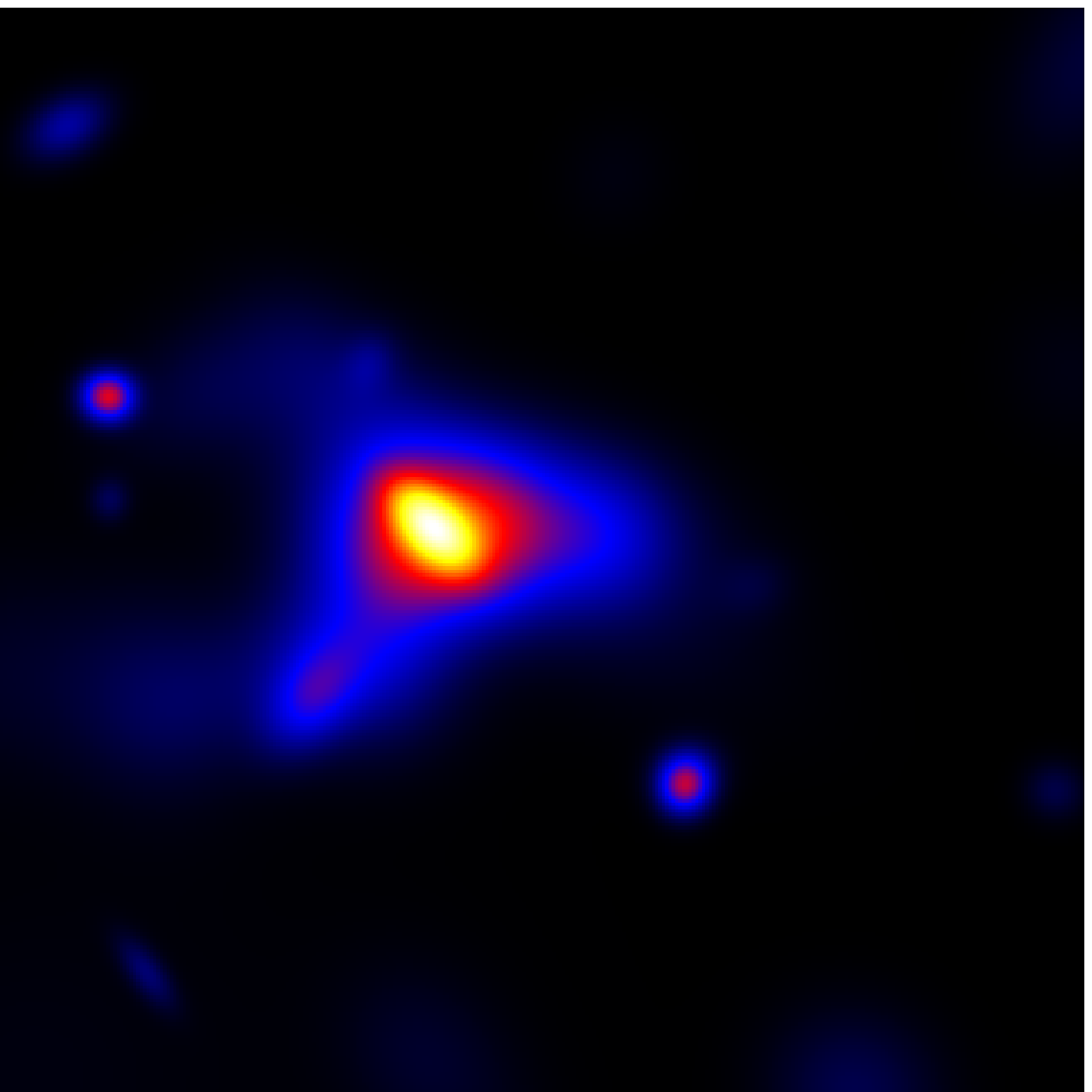} &
\includegraphics[width=0.176\textwidth]{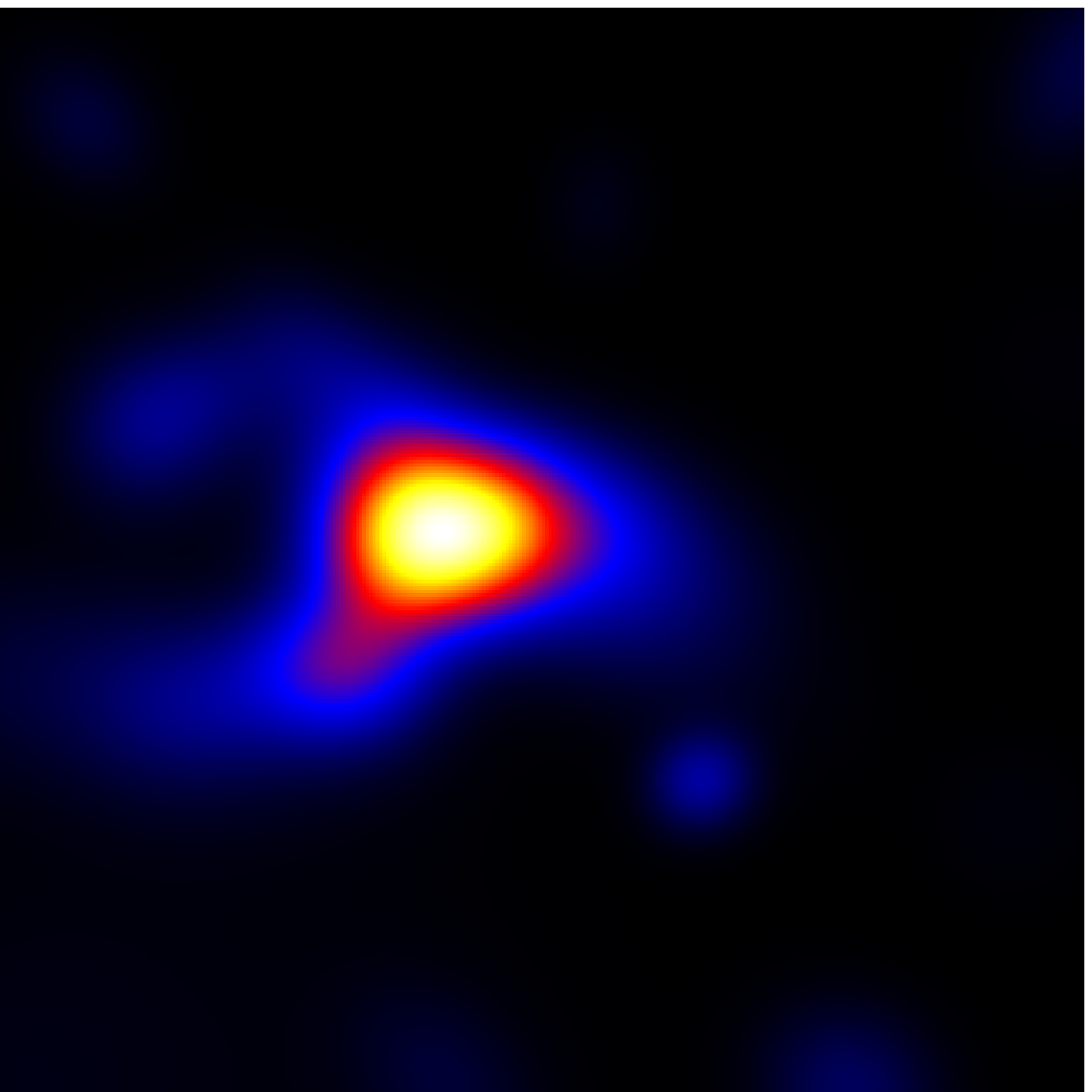} &
\includegraphics[width=0.176\textwidth]{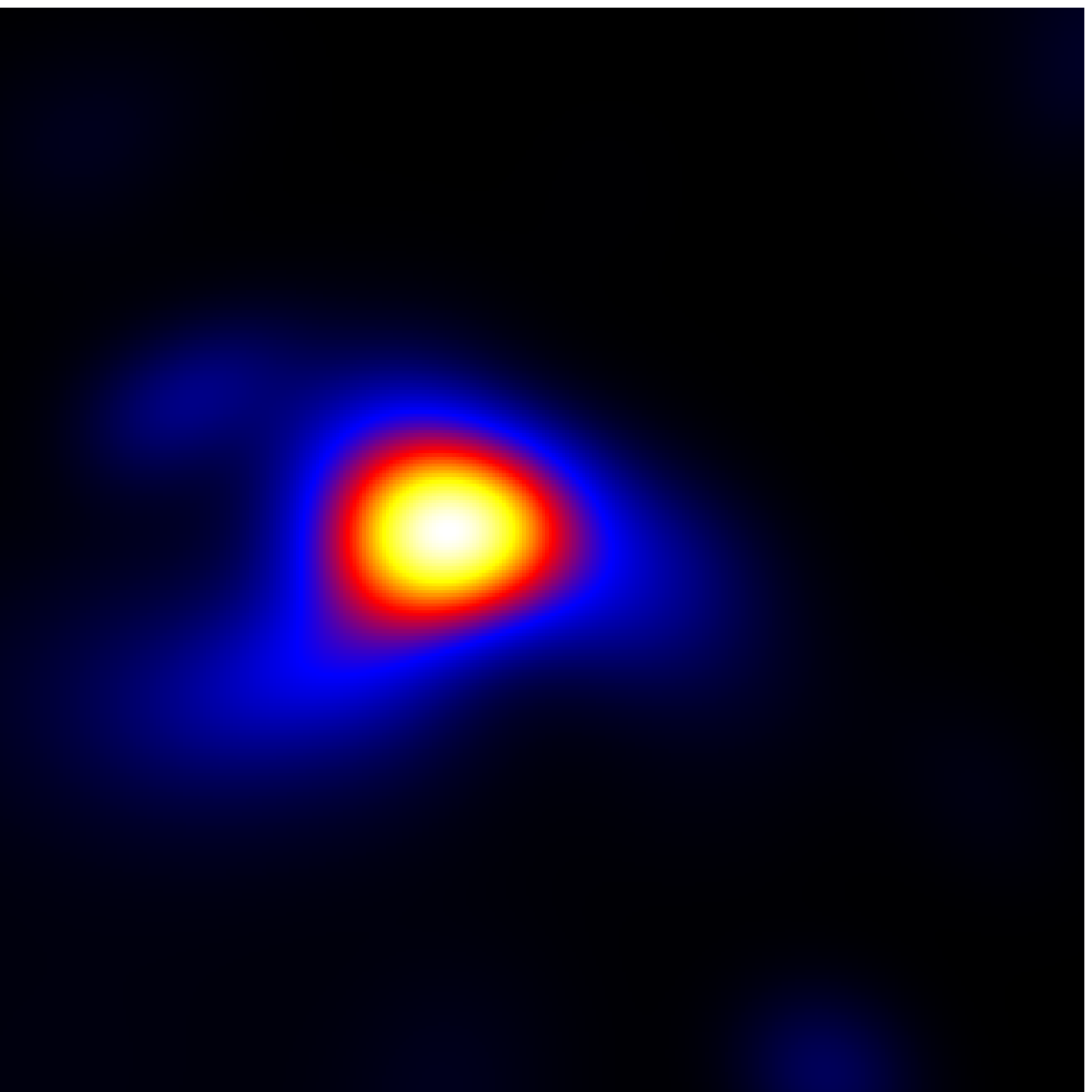} &
\includegraphics[width=0.176\textwidth]{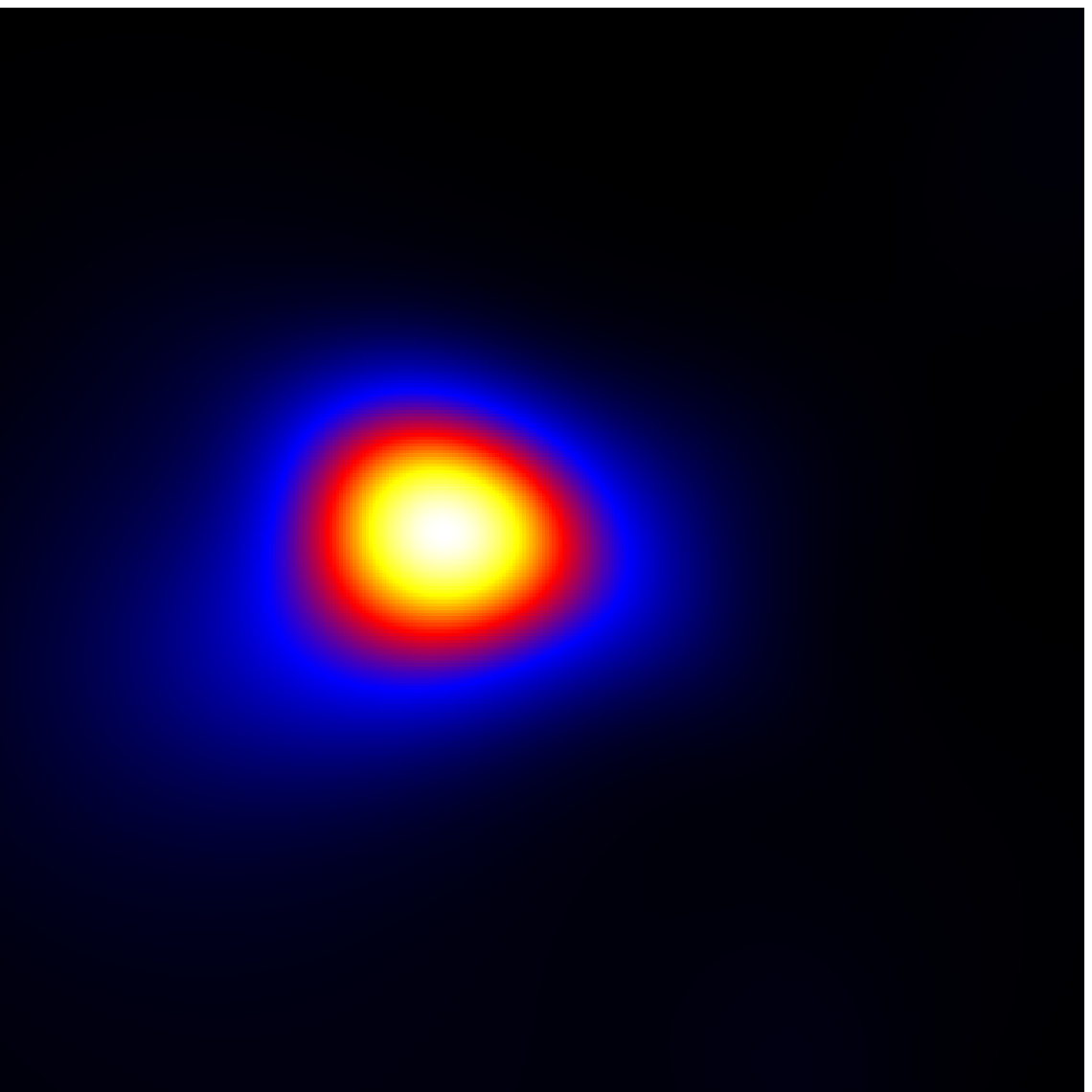} \\
$\lambda=0.1$ & $\lambda=0.3$ & \textbf{$\lambda=0.5$} & $\lambda=0.7$ & $\lambda=0.9$ \\
& & (default) & & \\
\end{tabular}
\caption{Reconstructions of a flare source observed with \emph{RHESSI} using different regularization strengths to control the sparseness of the reconstruction. Left ($\lambda=0.1$): Reconstructions with less regularization tend to over-resolve structures. Right ($\lambda=0.9$): Stronger regularization produces blob-like shapes with no apparent detail. In this paper $\lambda=0.5$ \revrem{was}\revadd{is} used for all reconstructions.}
\label{fig_lambda}
\end{figure*}

\revadd{The $\ell_{0}$ norm in the regularization term makes this an intractable combinatorial optimization problem. For this reason, we use the $\ell_{1}$ norm as a substitute, which makes this a convex optimization problem. The rationale and the impact of this common substitution are discussed in \citet{candes2008enhancing}.}

The sparseness parameter $\lambda$ is the only free parameter of the algorithm. It controls the regularization strength, to bias the solution towards more or less sparse reconstructions, as shown in Figure~\ref{fig_lambda}. Formulations of this kind are known as LASSO in statistics and machine learning \citep{Tib96}. \revadd{We use $\lambda=0.5$ for all reconstructions in this paper.} \revrem{For physical reasons}\revadd{Because the observed X-ray flux is intrinsically positive}, we impose positivity constraints on the image $\mathbfit{\hat{x}}$. \revadd{As all basis functions are non-negative, it suffices to add a non-negativity constraint for each basis coefficient}. With these constraints the final formulation of our algorithm VIS\_CS looks like this:


\begin{equation}
\begin{aligned}
& \underset{\mathbfit{\hat{x}}}{\text{minimize}}
& & (1-\lambda)\left\lVert \mathbf{A}\mathbfit{\hat{x}}-\mathbfit{y} \right\rVert_2^2 + \lambda\|\mathbf{D}\mathbfit{\hat{x}}\|_1 \\
& \text{subject to} & & \forall{\mathbfit{\hat{x}}_i} \geqslant 0,\\
\end{aligned}
\end{equation}
Thus VIS\_CS is a convex, linearly constrained quadratic optimization problem. Convex optimization problems are easier to solve than most other optimization problems: By construction, the only local optimum is also the global optimum \citep{ponstein1967seven}. There are many methods to solve convex, linearly constrained quadratic optimization problems \citep{boyd2011distributed,wolfe1959simplex,nesterov1994interior,kozlov1980polynomial}. We use Coordinate Descent with the Active Set heuristic to solve the optimization problem \citep{friedman2010regularization}. Coordinate Descent reconstructs images in reasonably short times, as the wall-clock comparison in Table~\ref{tbl_performance} shows. The VIS\_CS reconstruction time is largely independent of the number of reconstructed pixels, as the optimization problem is stated in the visibility domain and the sparse basis. On the other hand, complex sources typically require more time to reconstruct than simpler, sparse sources.

\begin{deluxetable*}{lr}
\tablecaption{Average wall-clock time to reconstruct a single $131 \times 131$ pixel image}
\tablehead{\colhead{Algorithm}&\colhead{Seconds}}
\startdata
uv\_smooth                   &  8.8 \\
Back projection              &  9.9 \\
\textbf{VIS\_CS} & \textbf{10}.\textbf{0} \\
Clean                        &  12.8 \\
Pixon                        &  3665.6 \\
\enddata
\tablecomments{Reported are the average reconstruction duration for an image cube with 15 time- and 19 energy-bins of the 19 July 2012 (04:34:00 - 05:26:04 UT) event with an Intel Xeon E5-2650 2.3GHz CPU. Due to time constraints, we timed Pixon only for a single \revrem{reconstruction}\revadd{time- and energy-bin}.}
\label{tbl_performance}
\end{deluxetable*}

\subsection{Gaussian distributions as sparse basis for X-ray source distributions}

\begin{figure*}
\begin{tabular}{ccccc}
Spatial (Pixel) basis & Total Variation basis & Gaussian basis & Ground truth \\
\includegraphics[width=0.23\textwidth]{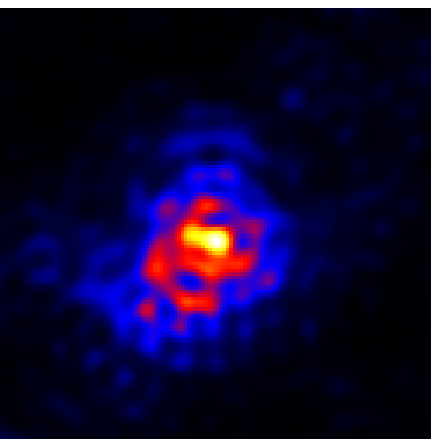} &
\includegraphics[width=0.23\textwidth]{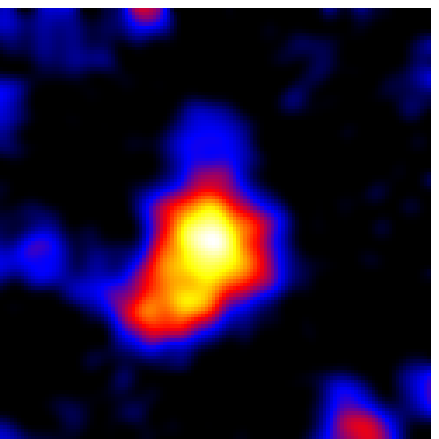} &
\includegraphics[width=0.23\textwidth]{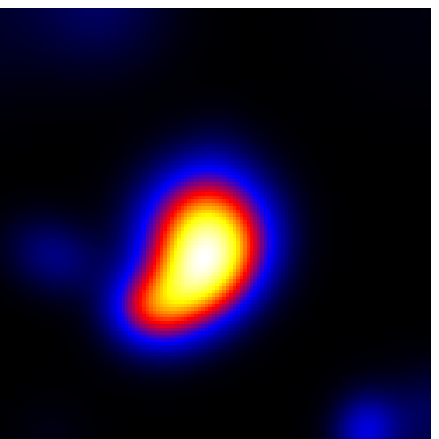} &
\includegraphics[width=0.23\textwidth]{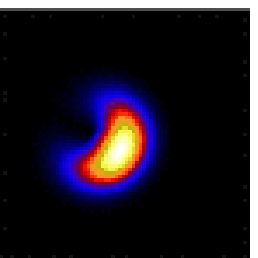} \\
\end{tabular}
\caption{Reconstructions of synthetic HESPE test data \citep{hespeit} using different bases compared with the ground truth (right image). Commonly used bases, such as Total Variation, yield poor results with this data. Our custom Gaussian basis (third image) yields a reconstruction that is closest to the ground truth.}
\label{fig_bases}
\end{figure*}

\begin{figure*}
\begin{tabular}{c}
\includegraphics[width=0.135\textwidth]{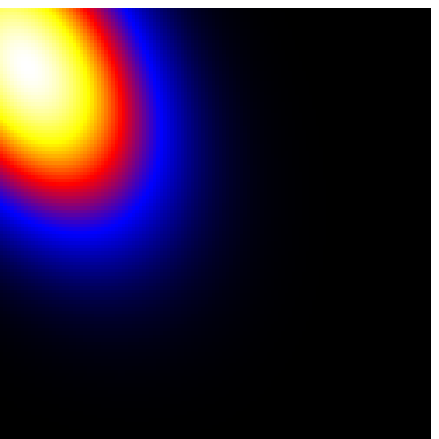}
\includegraphics[width=0.135\textwidth]{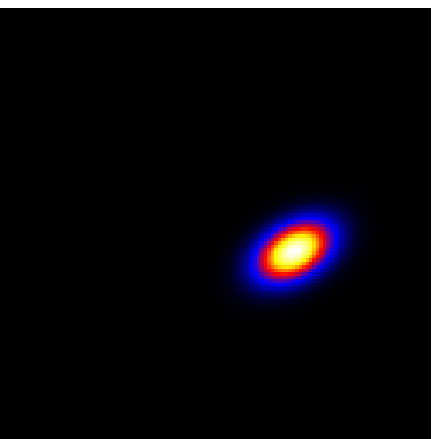}
\includegraphics[width=0.135\textwidth]{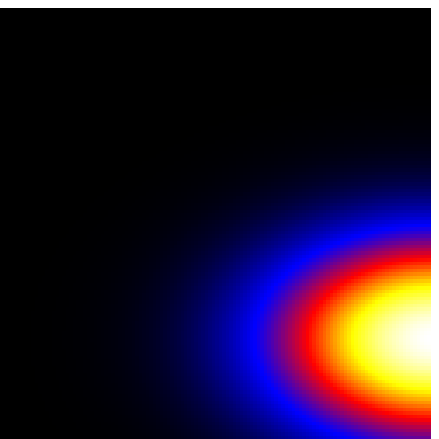}
\includegraphics[width=0.135\textwidth]{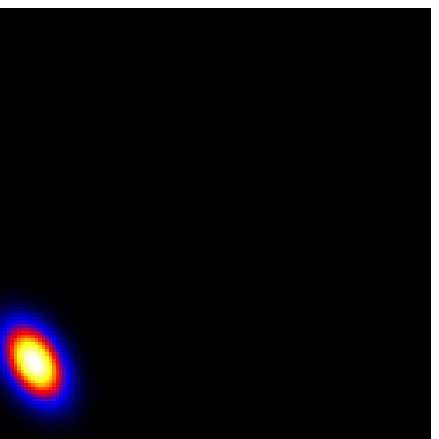}
\includegraphics[width=0.135\textwidth]{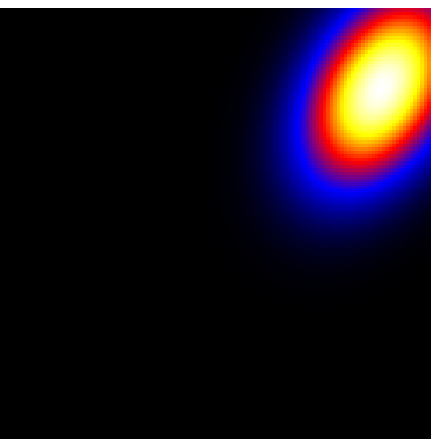}
\includegraphics[width=0.135\textwidth]{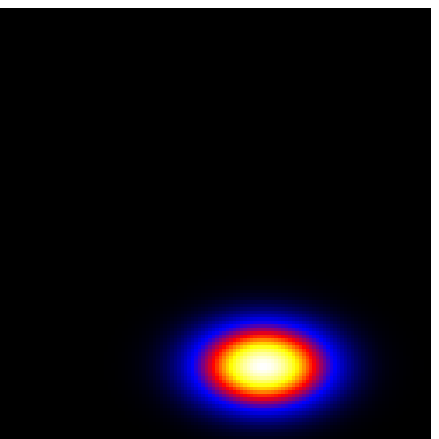}
\includegraphics[width=0.135\textwidth]{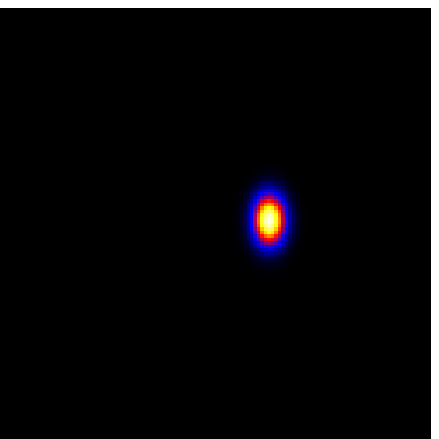}\\[0.2em]
\includegraphics[width=0.135\textwidth]{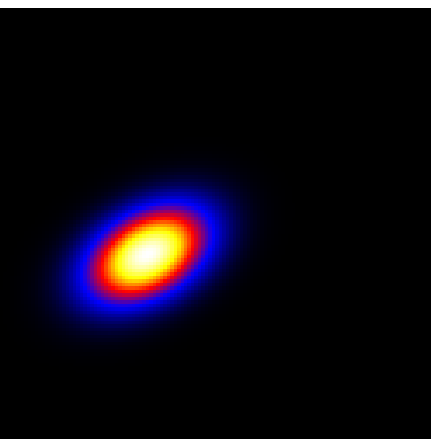}
\includegraphics[width=0.135\textwidth]{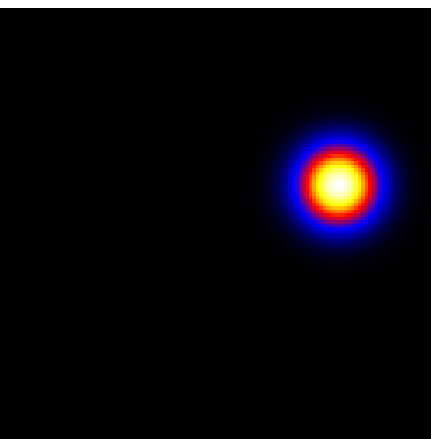}
\includegraphics[width=0.135\textwidth]{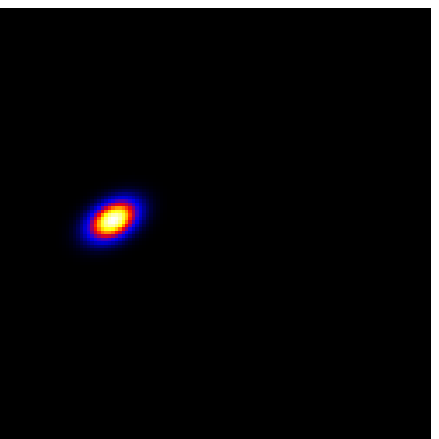}
\includegraphics[width=0.135\textwidth]{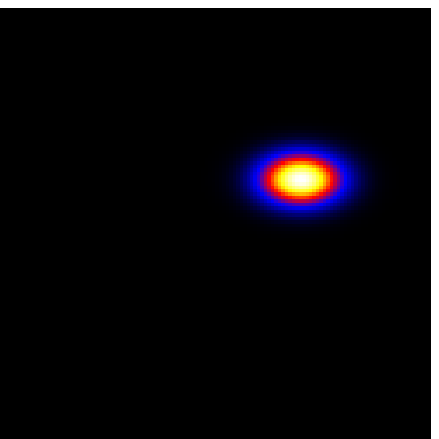}
\includegraphics[width=0.135\textwidth]{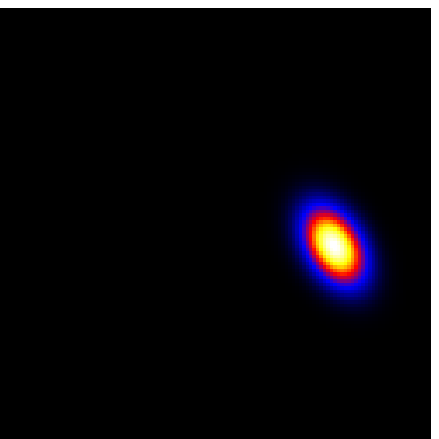}
\includegraphics[width=0.135\textwidth]{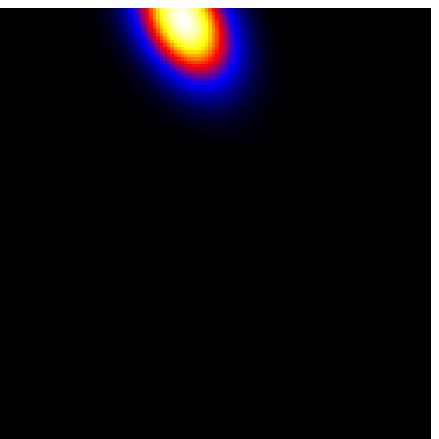}
\includegraphics[width=0.135\textwidth]{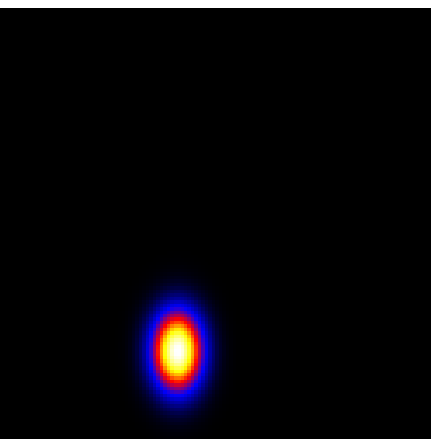}
\end{tabular}
\caption{Visualization of a few typical Gaussian distributions that constitute the basis functions of the Gaussian basis. The method assumes that X-ray source distributions can be represented by a sparse linear combination of such basis functions.}
\label{fig_basis_functions}
\end{figure*}

The success of Compressed Sensing hinges on the choice of the sparse basis $\mathbf{D}$. \citet{candes2006near} describe the successful use of a total-variation basis and a spatial (Pixel) basis to process natural images. \citet{starck2010sparse} experimented with Wavelets, Starlets, and Curvelets as sparse bases. In our experiments, we found that a custom Gaussian basis results in better reconstructions from \emph{RHESSI} data. Reconstructions with some common bases are shown in Figure~\ref{fig_bases}.


We use a custom Gaussian basis on the assumption that sources can be represented as linear combinations of a number of Gaussian distributions. This Gaussian basis is an overcomplete basis consisting of two-dimensional Gaussian distributions of various sizes, orientations, and locations in the $xy$-plane. Figure~\ref{fig_basis_functions} shows examples of Gaussian basis functions, which make up the chosen Gaussian basis. We \revrem{use}\revadd{randomly sample} up to $10^6$ basis functions from the infinite set of Gaussian distributions. The number of basis functions represents a trade-off between the reconstruction quality and speed. For practical reasons, we reduce the problem size by using fewer basis functions when reconstructing from more visibilities and vice versa. We find empirically that the reconstruction quality does not improve noticeably when using more than $10^6$ basis functions.

We use Gaussian distributions with a size of $1.5" \leqslant \sigma_{x,y}\leqslant 40"$, corresponding to typical observed solar flare source sizes. The eccentricity is limited by imposing $\sigma_x \leqslant 2.5\sigma_y \leqslant 2.5^2 \sigma_x$. Strongly eccentric X-ray sources (e.g. flare ribbons or loops) can still be reconstructed by superposing multiple smaller Gaussians. While we developed this set of basis functions specifically for X-ray flare reconstructions from \emph{RHESSI} measurements, the algorithm itself is not tied in any way to this specific sparse basis.

\section{Application on synthetic and real data} \label{sec:test}

\subsection{Tests with synthetic data}
The quality of an image reconstruction algorithm must be assessed in several dimensions. A good reconstruction algorithm produces accurate photometry, source locations, sizes, and morphologies. Ideally, those properties hold in cases with single and multiple sources of varying dynamic range, with high and low detector counts. In the following sections we systematically create and analyze synthetic scenarios to test VIS\_CS along those dimensions using specialized simulation software (part of the \emph{RHESSI} data analysis package in Solar Soft), which allows for three different detector count statistics, i.e. number of simulated counts per detector to emulate flaring sources of different intensities. We use 136 visibilities from \emph{RHESSI} detectors 3 to 9. All algorithms are used with their default parameters \footnote{\revadd{The default parameters of all \emph{RHESSI} imaging algorithms are documented at \url{https://hesperia.gsfc.nasa.gov/ssw/hessi/doc/params/hsi_params_image.htm}}}. These defaults \revrem{were carefully chosen}\revadd{were empirically optimized by the respective algorithm experts} to yield \revrem{acceptable}\revadd{reasonable} \revrem{images}\revadd{results} for typical flare sources, but \revrem{the best quality image can often only be constructed by systematically changing these defaults}\revadd{often the results can be improved by changing parameters on a case by case basis. However, this requires} in-depth knowledge of the individual algorithms. Many casual \emph{RHESSI} data users are not familiar with the intricacies of all the different algorithms and for certain applications like imaging spectroscopy it is impractical to fine-tune the parameters for individual images. Hence for our tests we generally use default parameters, unless otherwise specified.

\subsubsection{Dynamic range and photometry}
\begin{figure*}
\begin{tabular}{cccc}
1:1 & 1:5 & 1:10 & 1:50 \\
\includegraphics[width=0.23\textwidth,trim={10pt 20pt 20pt 30pt,clip}]{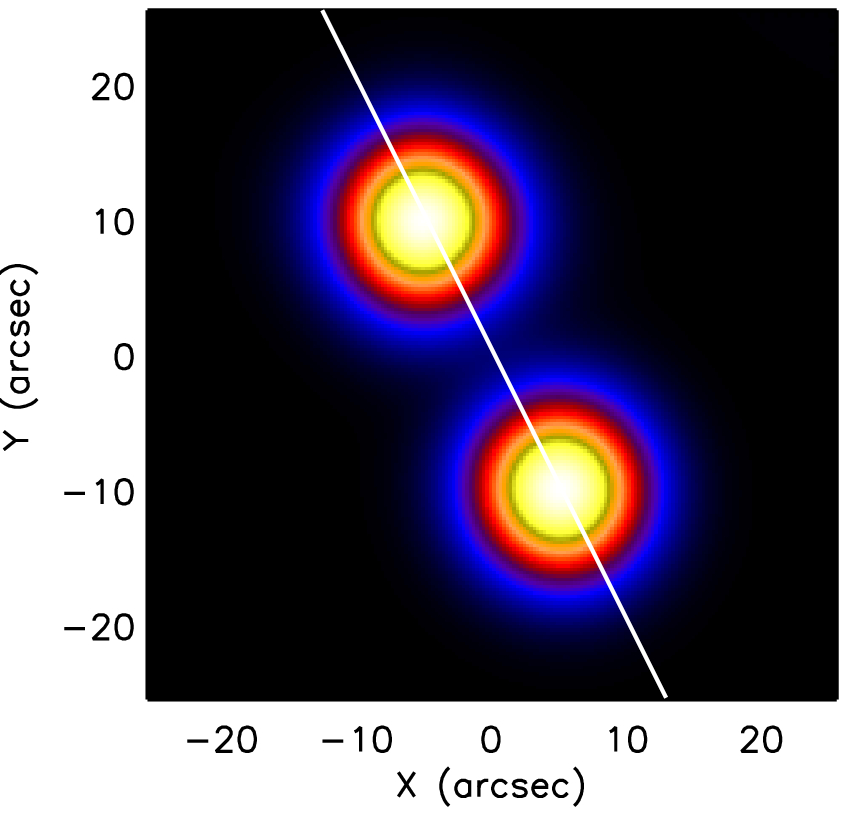} &
\includegraphics[width=0.23\textwidth,trim={10pt 20pt 20pt 30pt,clip}]{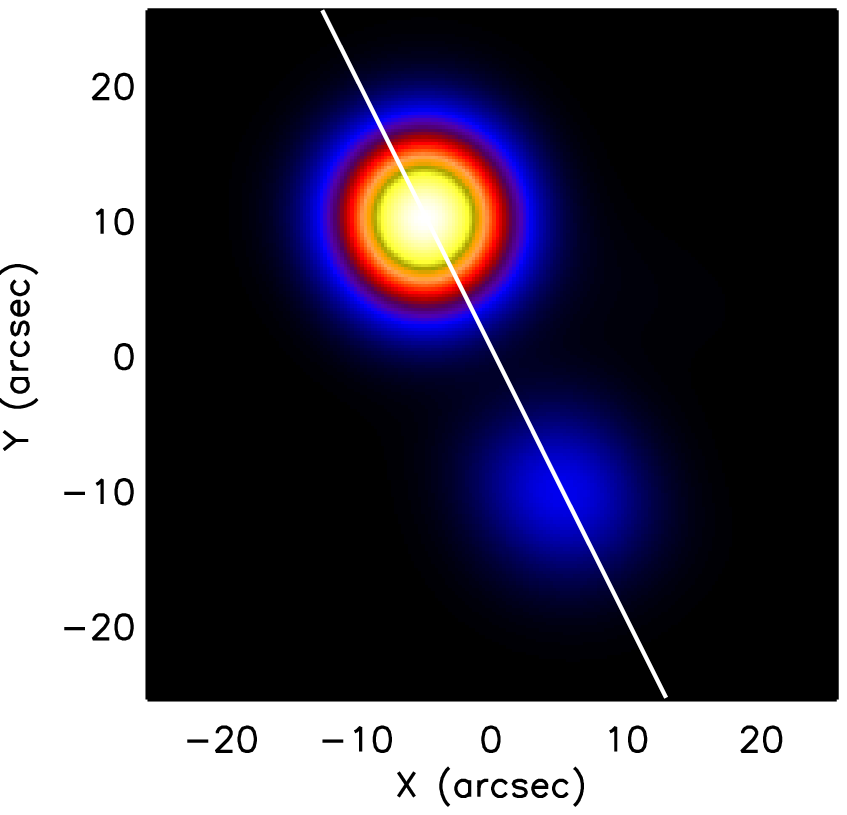} &
\includegraphics[width=0.23\textwidth,trim={10pt 20pt 20pt 30pt,clip}]{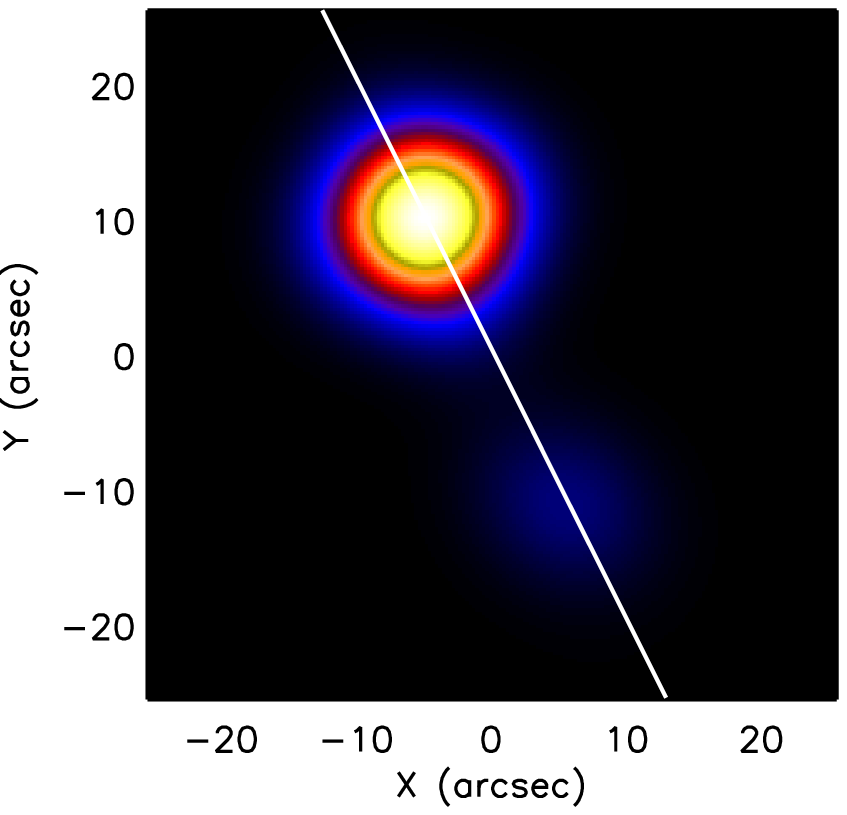} &
\includegraphics[width=0.23\textwidth,trim={10pt 20pt 20pt 30pt,clip}]{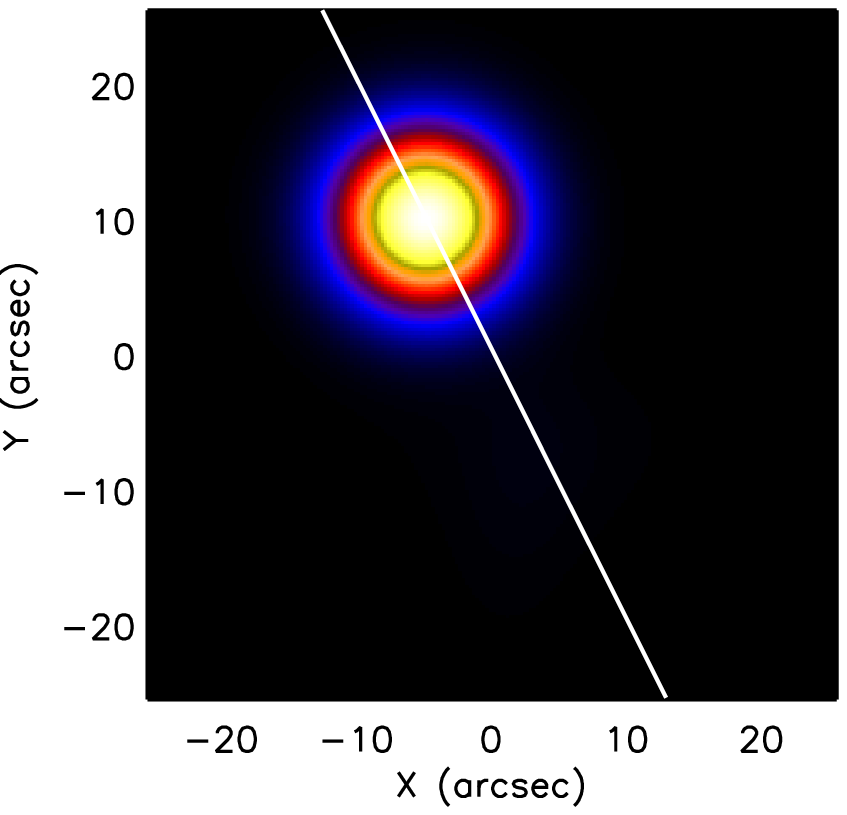}\\
\includegraphics[width=0.23\textwidth,trim={30pt 60pt 0pt 0pt,clip}]{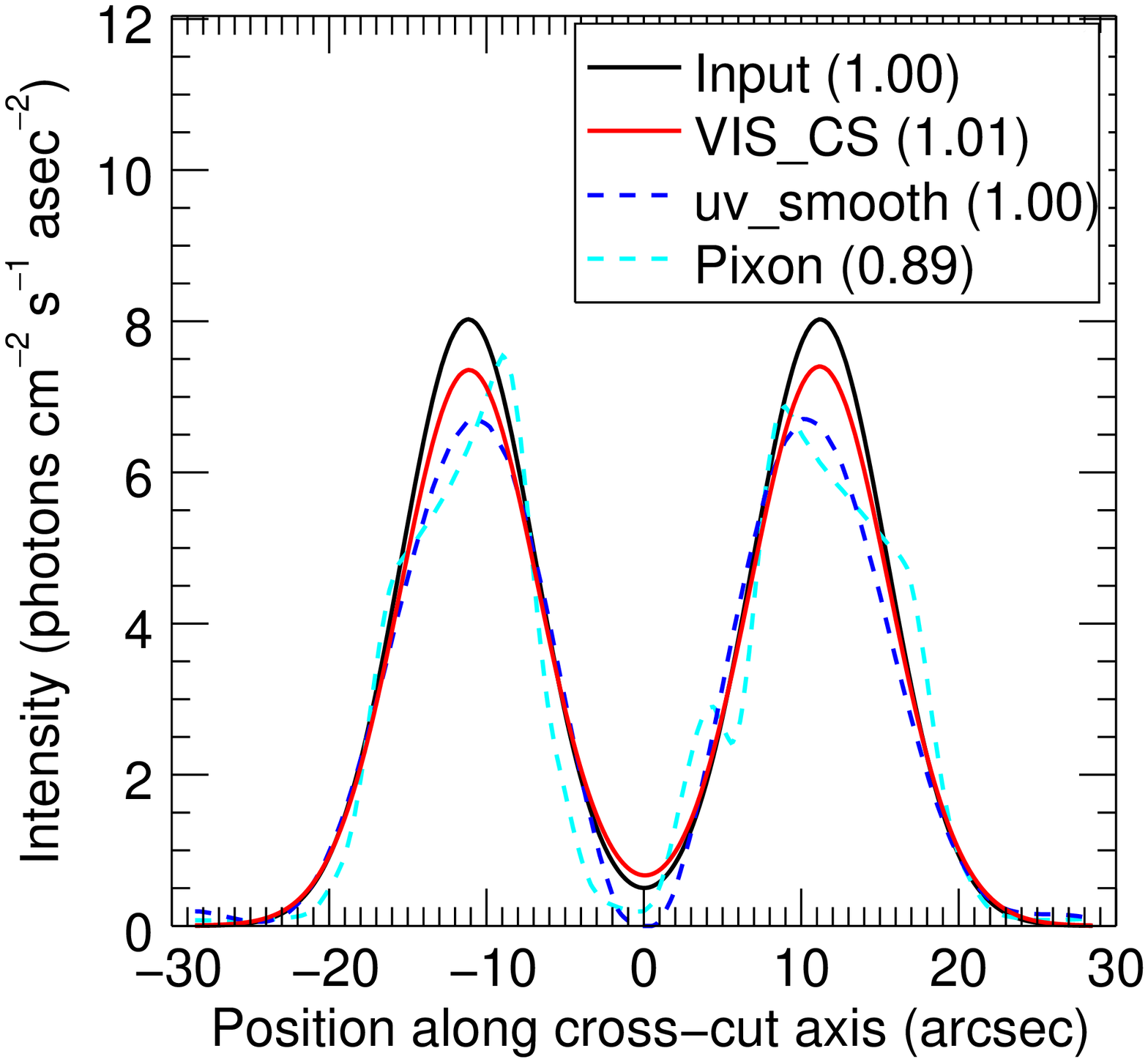} &
\includegraphics[width=0.23\textwidth,trim={30pt 60pt 0pt 0pt,clip}]{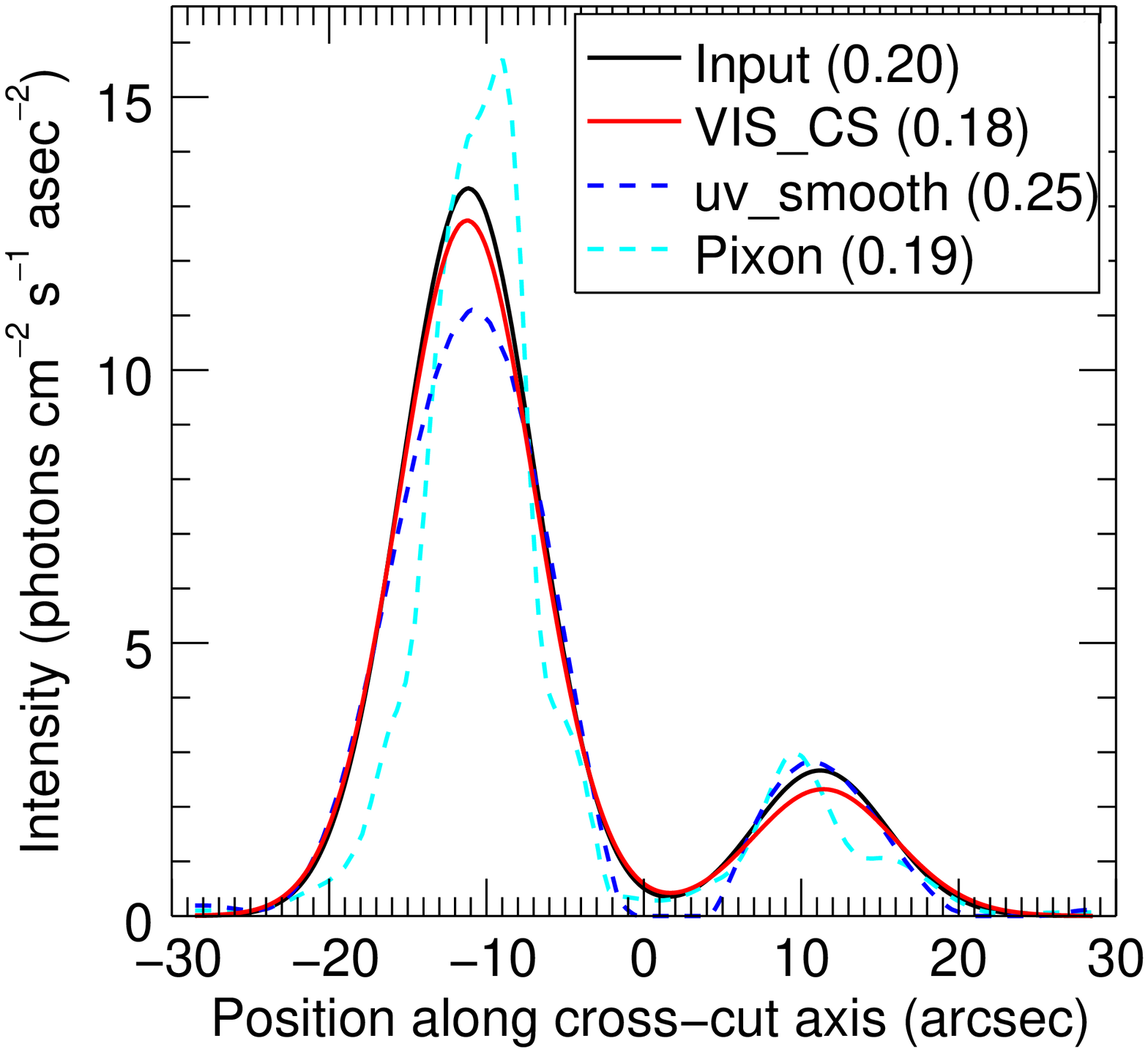} &
\includegraphics[width=0.23\textwidth,trim={30pt 60pt 0pt 0pt,clip}]{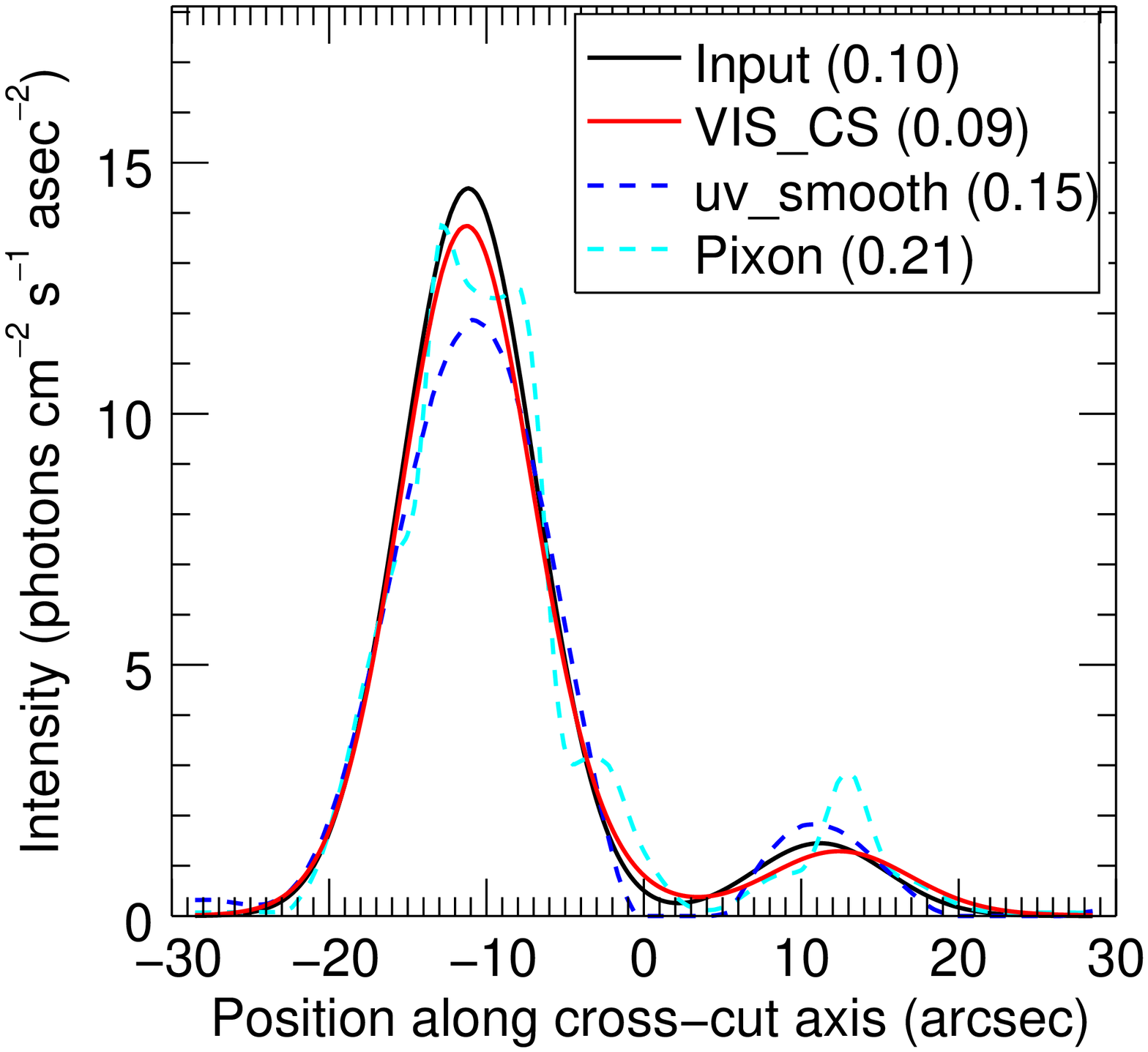} &
\includegraphics[width=0.23\textwidth,trim={30pt 60pt 0pt 0pt,clip}]{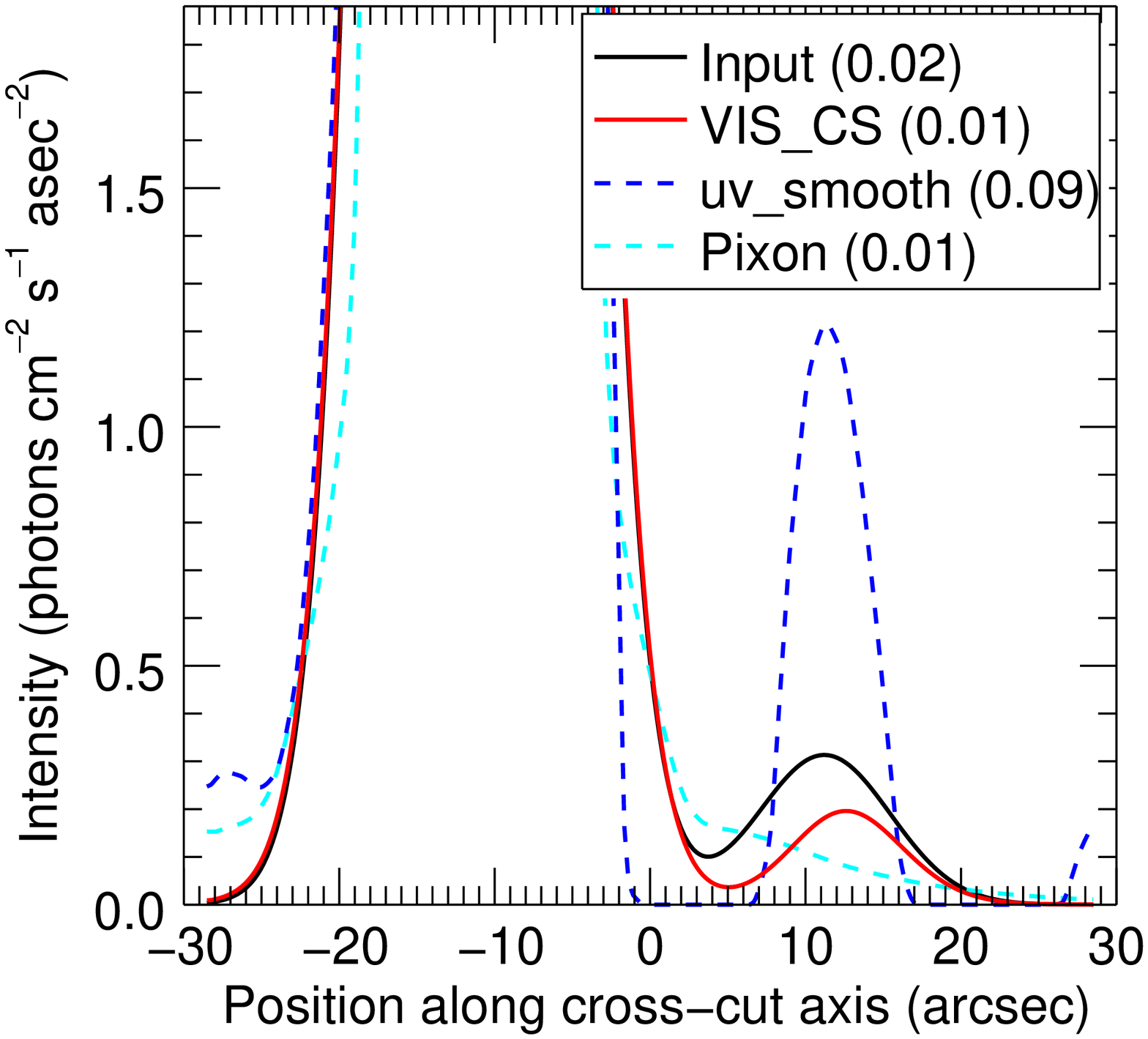}
\end{tabular}
\caption{Top row: VIS\_CS reconstructions of two circular Gaussian sources (FWHM = $10"$), separated by 22", with different peak flux ratios. The simulations use peak flux ratios of 1:1, 1:5, 1:10 and 1:50 per column. Bottom row: Intensity profiles along the white cross-cut axis. The reconstructed peak ratios are shown in parentheses after each algorithm.}
\label{fig_dynamic_range}
\end{figure*}

To test the dynamic range of the algorithms we recreate a test proposed by \citet{alM09}, where two circular Gaussian sources with a FWHM of 10" and varying peak ratios are simulated. The sources are separated by 22". We use simulations with 100'000 cts det$^{-1}$ emulating hight counting statistics. The results in Figure~\ref{fig_dynamic_range} show that VIS\_CS faithfully reproduces simple source morphologies and peak ratios. The top row displays VIS\_CS reconstructions, the bottom row shows one-dimensional cross cuts of reconstructions from several algorithms. In all four cases the VIS\_CS reconstructions have total fluxes between 98\% and 99.5\% of the true input fluxes. Therefore, VIS\_CS marginally underestimates the total flux of sources in this test. VIS\_CS reconstructs 100.6\% (1:1), 91.1\% (1:5), 93.8\% (1:10) and 61.5\% (1:50) of the true peak ratios. \revrem{These results indicate that VIS\_CS has no inherent dynamic range limits, but that dynamic range is limited by the noise in the simulated measurements.} With the noise statistics simulated in these tests VIS\_CS reliably identifies weak sources in cases with a dynamic range of 1:50. The ability of VIS\_CS to reproduce consistent source shapes is also noteworthy.

\begin{figure*}
\begin{tabular}{c}
\includegraphics[width=0.24\textwidth,trim={0px 10px 20px 0px},clip]{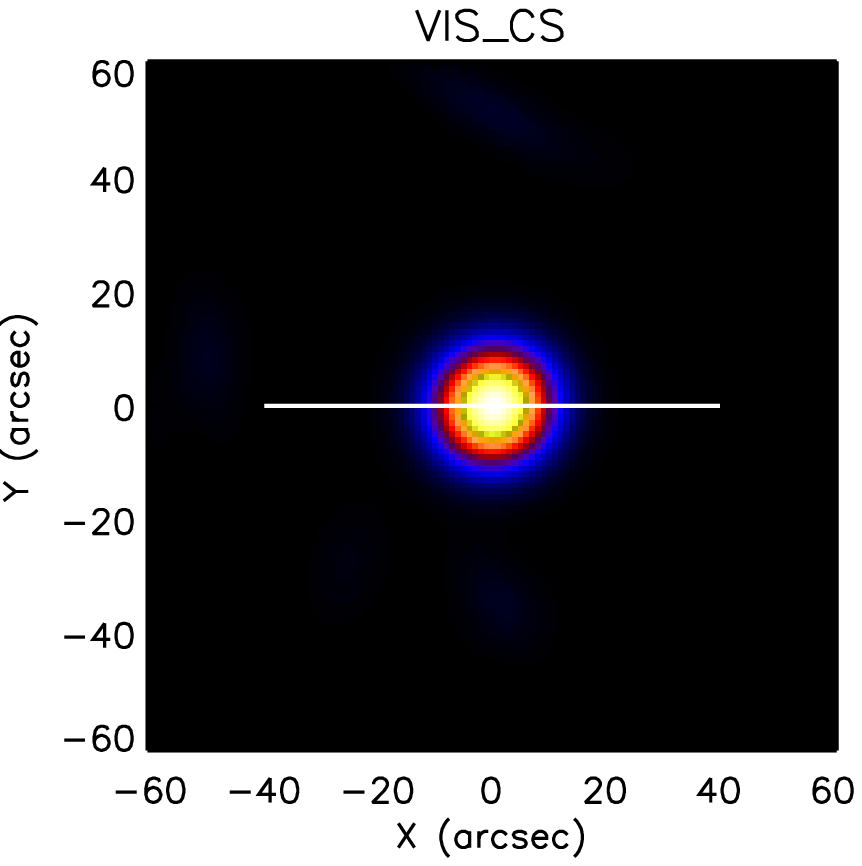}
\includegraphics[width=0.24\textwidth,trim={20px 20px 20px 20px},clip]{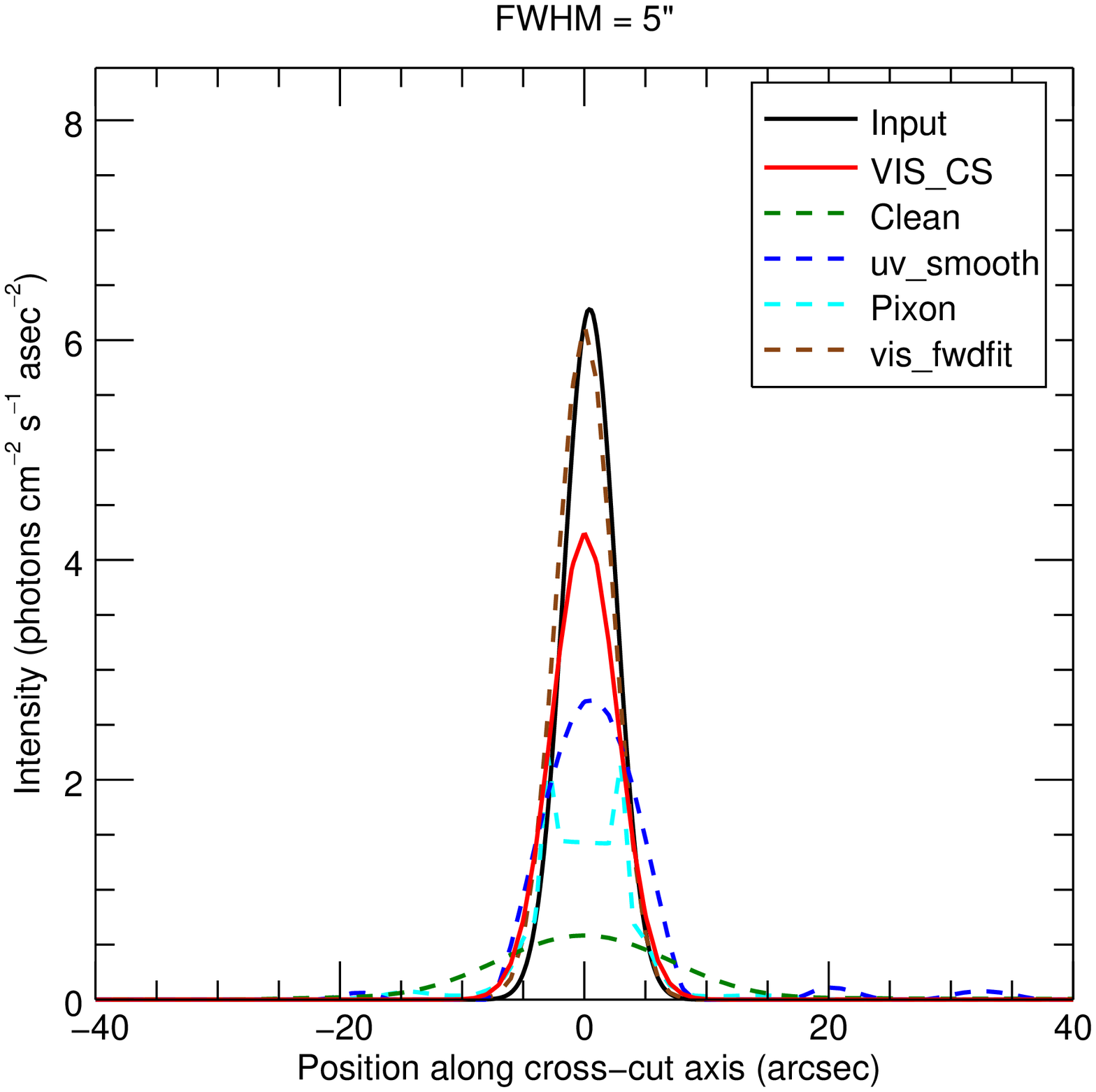}
\includegraphics[width=0.24\textwidth,trim={20px 20px 20px 20px},clip]{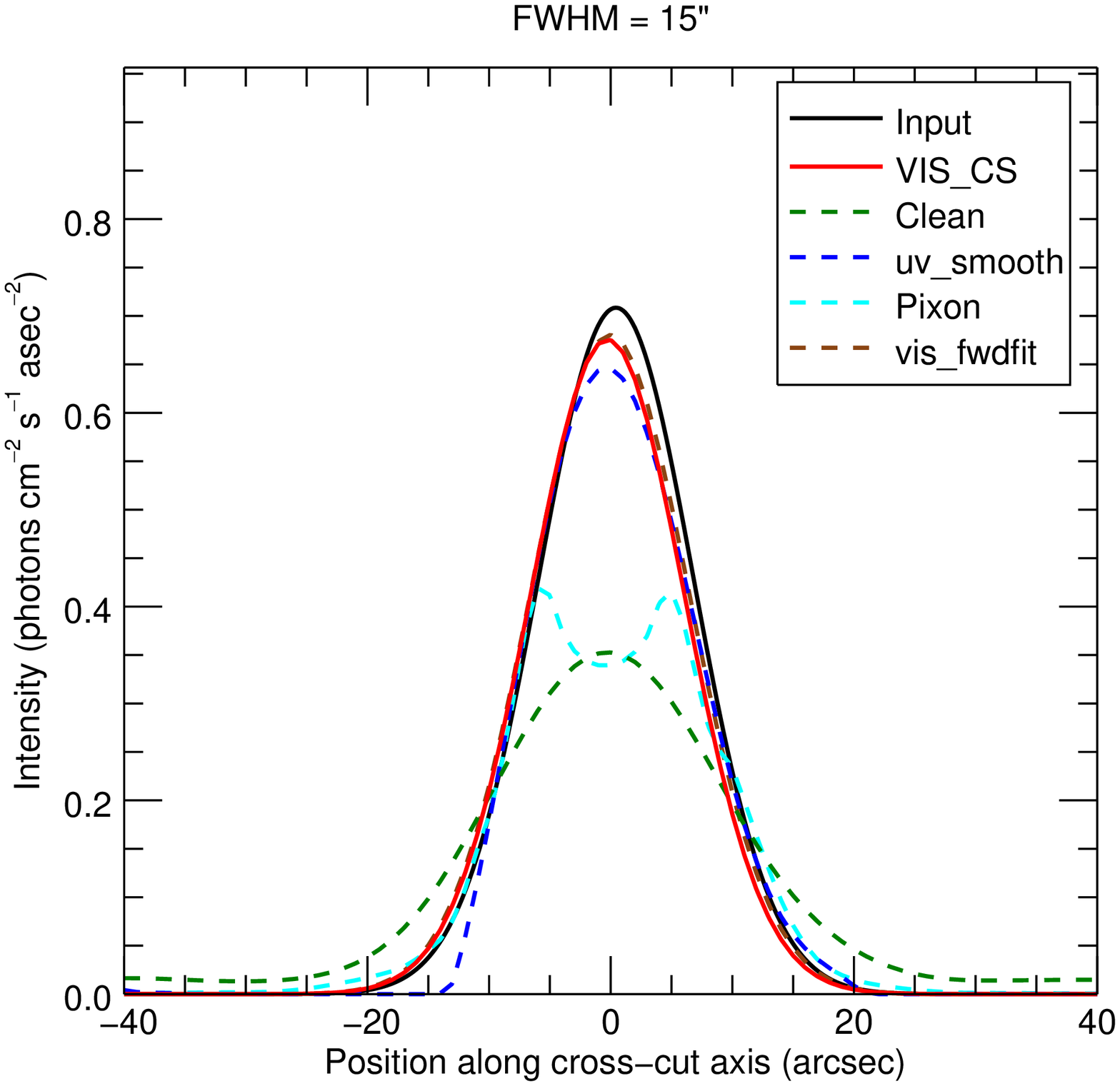}
\includegraphics[width=0.24\textwidth,trim={20px 20px 20px 20px},clip]{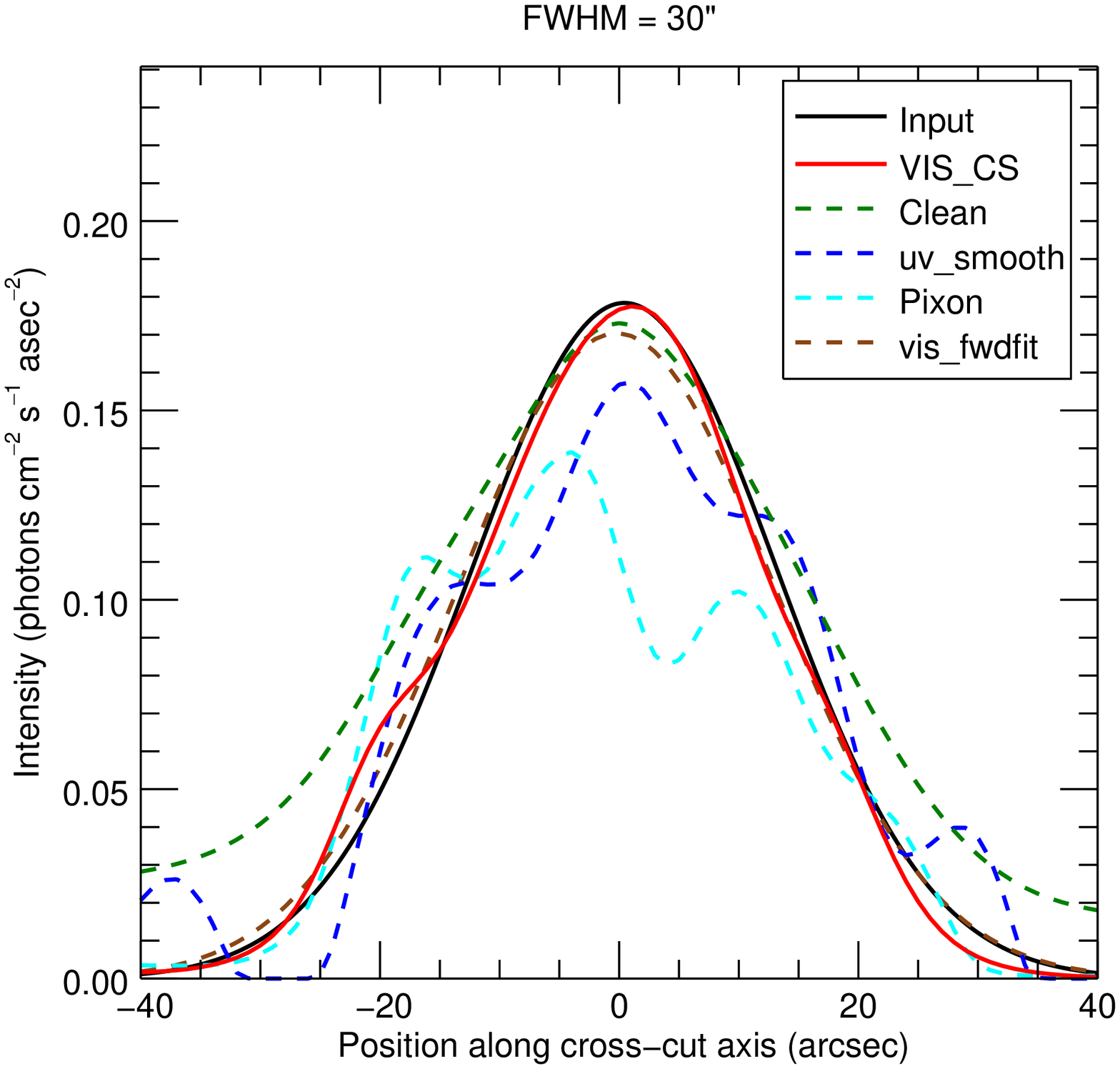}
\end{tabular}
\caption{Tests with a single circular Gaussian source with FWHM $\in \{ 5", 15", 30" \} $. The  image \revrem{in the upper left corner}\revadd{on the left} is a VIS\_CS reconstruction of the FWHM = 15" case, where the horizontal white line marks the cross-cut axis. The remaining images show intensity profiles along this horizontal cross-cut axis.}
\label{fig_source_size}
\end{figure*}

We test the absolute accuracy of source photon fluxes and source sizes reconstructed with VIS\_CS with a similar setup. We simulate a single circular Gaussian source in various sizes and plot horizontal cross-cuts through the center of the source. The horizontal axis was chosen arbitrarily, other axes yield similar results. This time, we use simulations with 10'000 cts det$^{-1}$ to see the impact of a reduced signal to noise ratio. Figure~\ref{fig_source_size} shows that VIS\_CS reconstructs total fluxes, peak fluxes, and source sizes with good accuracy. The only algorithm to produce better results is Forward Fit, which we configured to reconstruct a single circular Gaussian source. It is not surprising that small sources (FWHM $\leqslant 5" $) present a challenge to all reconstruction algorithms since the finest \revrem{grid}\revadd{subcollimator} used for this reconstruction has a pitch of 6.79". Sources of medium size (FWHM = 15") are handled better by all algorithms. Very large sources with a FWHM $\geqslant 30"$ pose more of a challenge, as these sources start to fragment in reconstructions. In this test with default parameters, Clean and VIS\_CS are the most resilient reconstruction algorithms. VIS\_CS reconstructs 105.2\%, 104.4\%, 102.6\% of the true total flux and 67.9\%, 97.7\%, 97.4\% of the true peak amplitudes. It reconstructs sources of accurate sizes and without side-lobes. It is apparent that Pixon and Clean require \revrem{source-specific} tuning \revadd{per test} for accurate results. Without tuning, Clean cannot reconstruct small sources and by design over-estimates source sizes. uv\_smooth overestimates the source fluxes (171.2\%, 125.2\%, 107.0\%) and tends to break up large sources. Without tuning Pixon shows over-resolution artifacts in all tests.

\subsubsection{Spatial source localization}
\begin{figure*}
\begin{tabular}{c}
\includegraphics[width=0.24\textwidth,trim={20px 20px 20px 20px},clip]{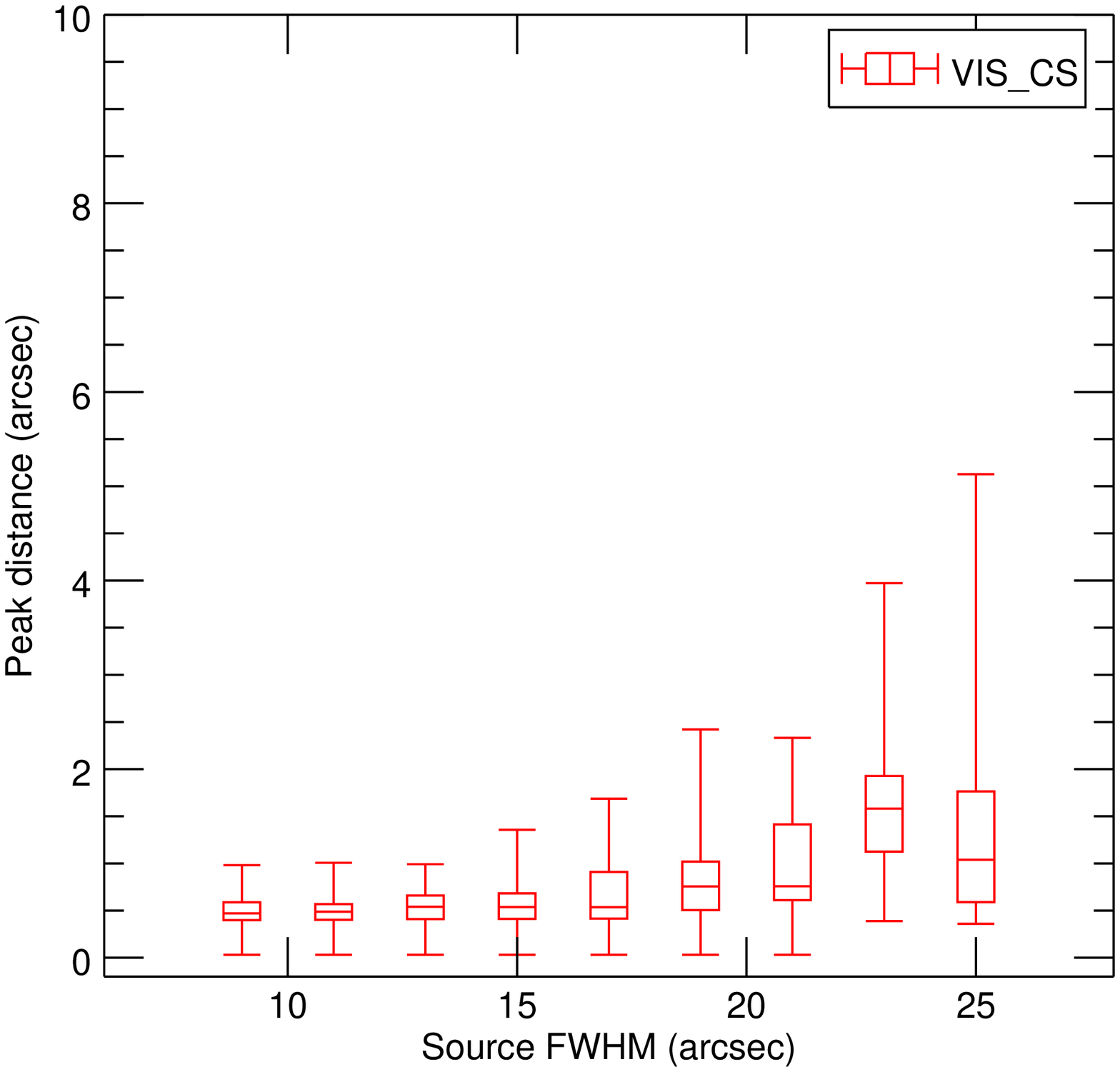}
\includegraphics[width=0.24\textwidth,trim={20px 20px 20px 20px},clip]{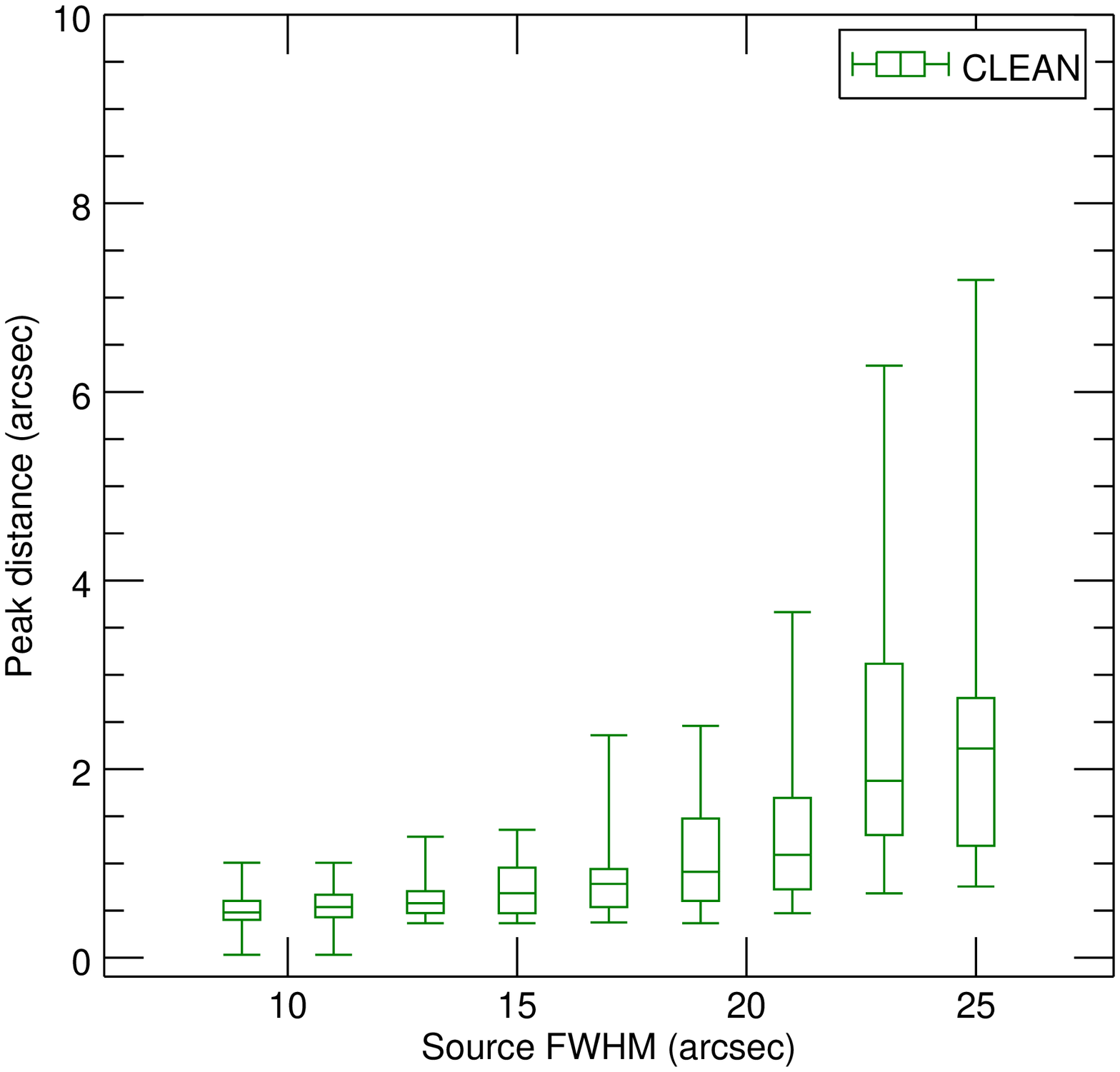}
\includegraphics[width=0.24\textwidth,trim={20px 20px 20px 20px},clip]{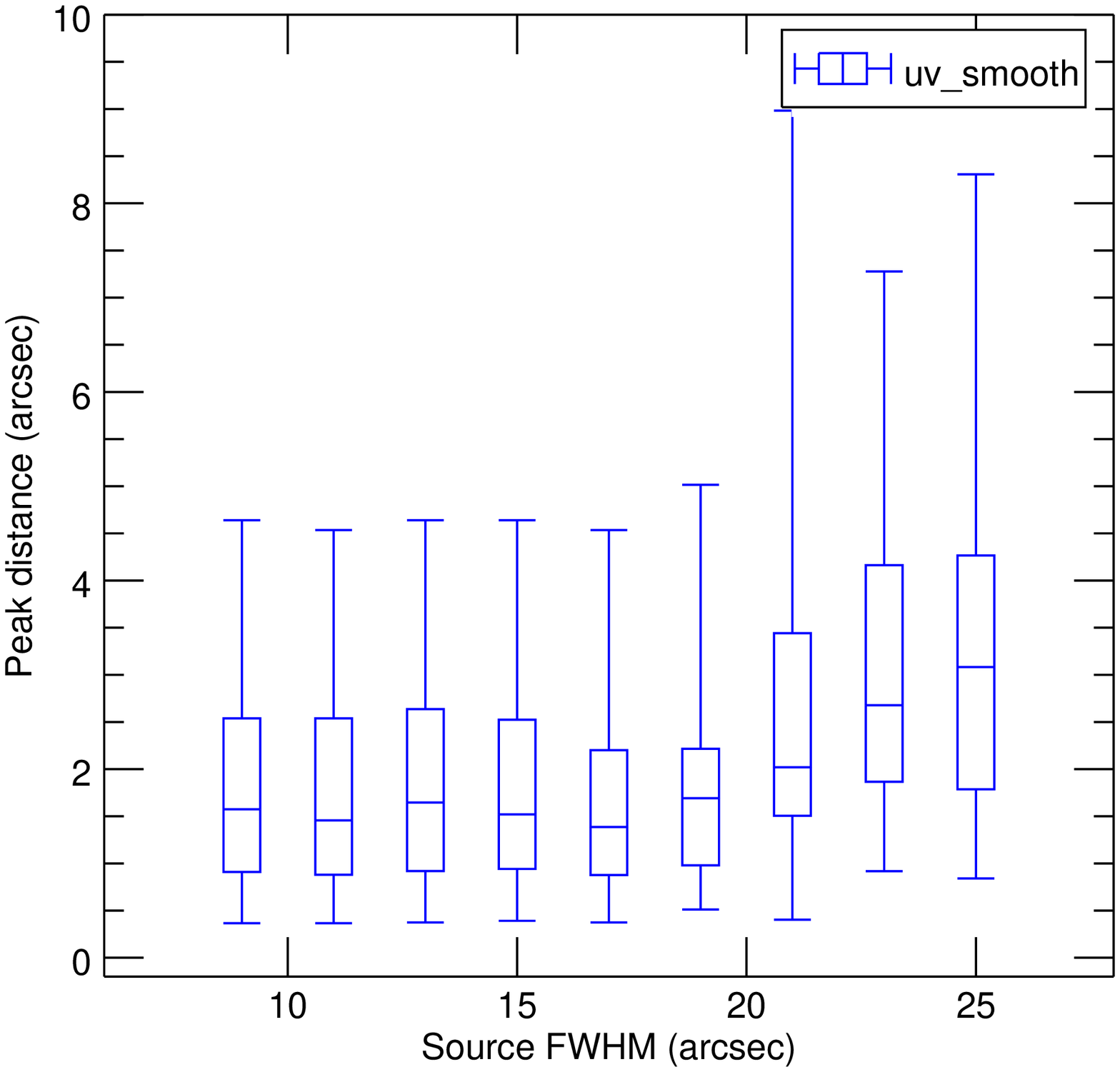}
\includegraphics[width=0.24\textwidth,trim={20px 20px 20px 20px},clip]{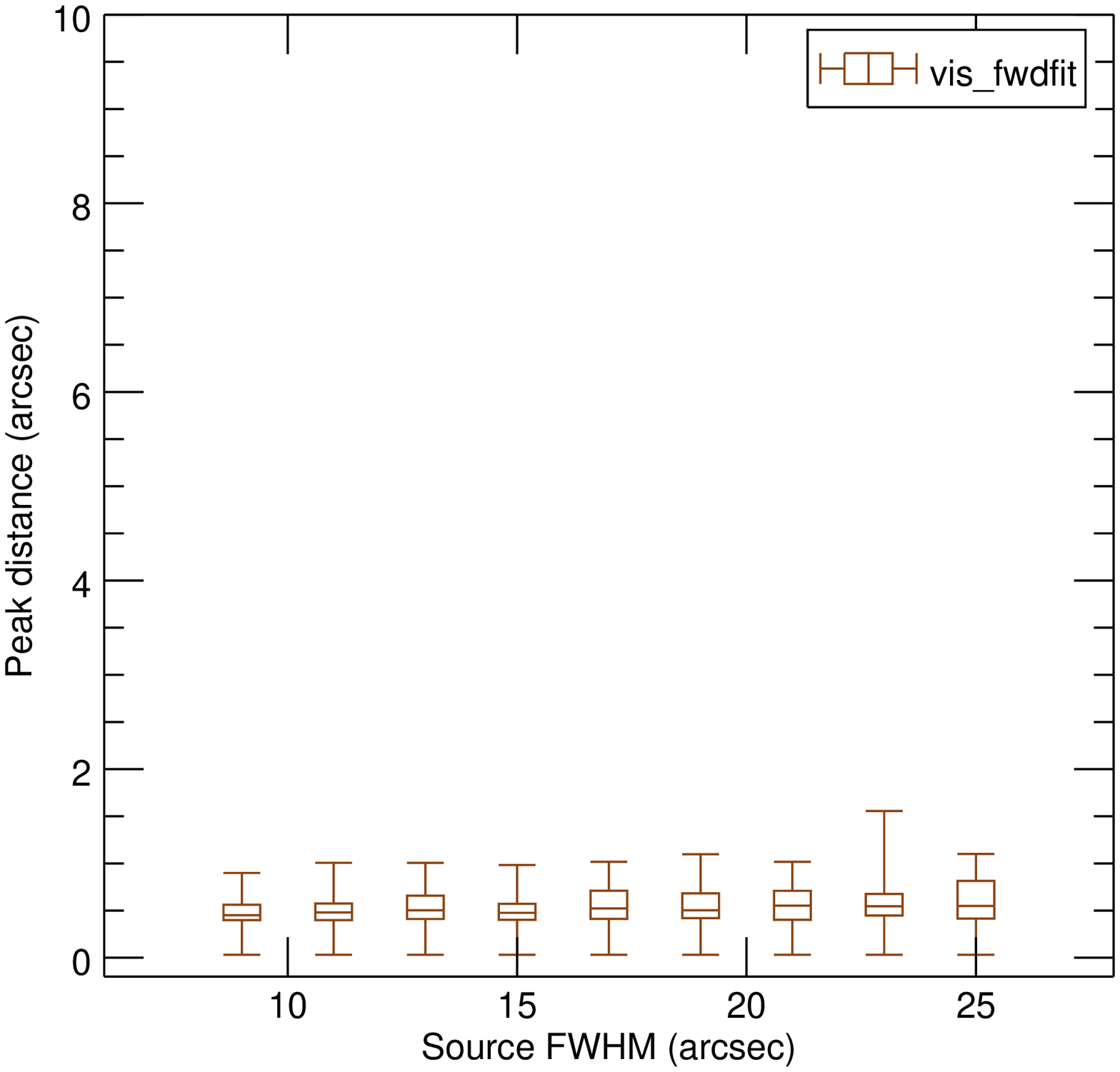}
\end{tabular}

\caption{Analysis of distances between reconstructed peak flux locations and actual peak flux locations of 1'500 simulated circular sources. Whiskers mark extremal distances, 25\% and 75\% distance percentiles are shown as boxes, the centerline represents median distances. Left to right, top to bottom: VIS\_CS, Clean, uv\_smooth, vis\_fwdfit. }

\label{fig_peakdist}
\end{figure*}

We analyze the ability of VIS\_CS to localize peak source locations. For this test we create 1\,500 instances with a single circular source and constant total flux at random locations. Figure~\ref{fig_peakdist} shows the distances of the reconstructed peak flux to the true center of the source. As expected, the model fitting algorithm vis\_fwdfit localizes sources with the highest accuracy of all tested algorithms. VIS\_CS reconstructs peak locations with similar accuracy up to FWHM $\geq 15"$. As we simulate sources with a constant total flux, larger simulated sources exhibit flatter peaks. Therefore, reconstruction errors affect the location of the maximum of larger sources more heavily.

\subsubsection{Performance for complex morphologies and low count rates}
\begin{figure*}
\includegraphics[width=0.95\textwidth,trim={0px 0px 60px 0px},clip]{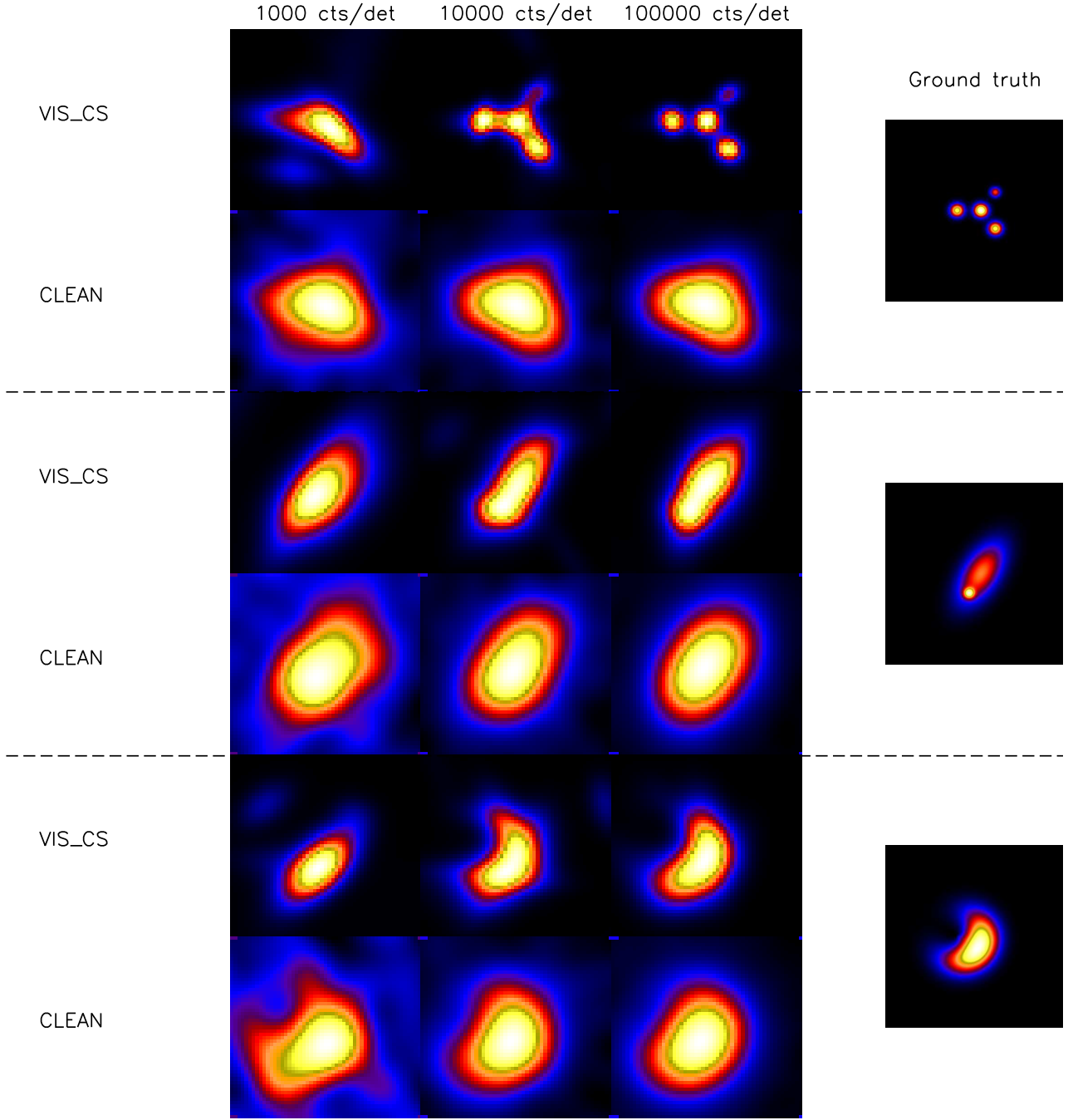}
\caption{Comparison of VIS\_CS and Clean reconstructions of synthetic simulations of different source shapes and count statistics. Left to right: Three different simulated detector count statistics (1'000 cts det$^{-1}$, 10'000 cts det$^{-1}$, 100'000 cts det$^{-1}$) and the artificial input map (i.e. ground truth). Top to bottom: Three different source morphologies (several small, compact sources, an elongated larger source, flare loop) inspired by real events from 23 July 2002 (00:29:10 - 00:30:19 UT, 36 - 41 keV), 2 December 2003 (22:54:00 - 22:58:00 UT, 18 - 20 keV), and 23 August 2005 (14:27:00 - 14:31:00 UT, 14 - 16 keV). Reconstructions \revadd{of this dataset with}\revrem{from} other algorithms are \revrem{available}\revadd{shown} in \citet{hespeit}.}
\label{fig_synthetic}
\end{figure*}

The above tests were performed using simple, circular Gaussian sources. To evaluate the performance of VIS\_CS qualitatively, with realistic, more complex morphologies, we use synthetic source configurations that were created as part of the EU FP7 HESPE project \citep{csillaghy2012mainstreaming}. These configurations were inspired by real events as observed by \emph{RHESSI}. We tested VIS\_CS with three configurations that represent more complex source morphologies; four compact sources, one compact source in the presence of an extend source, and a loop-shaped source. We compare VIS\_CS with Clean and the simulated input map (ground truth), using detectors 3 to 8 and the default parameter settings for Clean. For VIS\_CS, all detectors are used. Figure~\ref{fig_synthetic} shows the resulting maps for the three different detector count statistics and in comparison with the ground truth. According to \citet{hespeev} most existing algorithms benefit from parameter tuning, i.e. selection of detectors to be used and algorithm-specific parameter optimization such as number of iterations in the case of Clean. Here we deliberately neglected such optimization, other than detector selection, for both algorithms, and find that VIS\_CS reproduces the source morphology with the highest accuracy and no over-resolution artefacts, even in the case of low count rates.

\subsection{Application on observed X-ray flare data}
Solar flares are highly individual, with strongly varying morphology, depending on the photon energy and the time at which they are observed. One can broadly classify sources into three main groups:
\begin{enumerate}
\item Extended sources observed at low photon energies, usually seen in the solar corona, that originate from hot plasma;
\item Compact sources observed at higher photon energies originating from the chromosphere;
\item Loop-shaped sources originating from hot plasma that fills out magnetic loops connecting the corona with the chromosphere.
\end{enumerate}
Each of these sources provides different challenges for imaging algorithms. For example, \revrem{grids}\revadd{subcollimators} that correspond to a resolution finer than the source dimension will not lead to modulated flux and thus contribute only noise to the image reconstruction. The challenge for the user is thus to select the appropriate \revrem{grids}\revadd{subcollimators}, algorithm, and algorithm-specific parameters for a given source. We applied VIS\_CS on flares observed with \emph{RHESSI} that show these three typical morphologies and compare its performance with the reconstructions from the Clean and Pixon algorithms.

\subsubsection{Extended and compact sources}
\begin{figure*}
\includegraphics[width=0.98\textwidth,trim={0 0 10px 10px},clip]{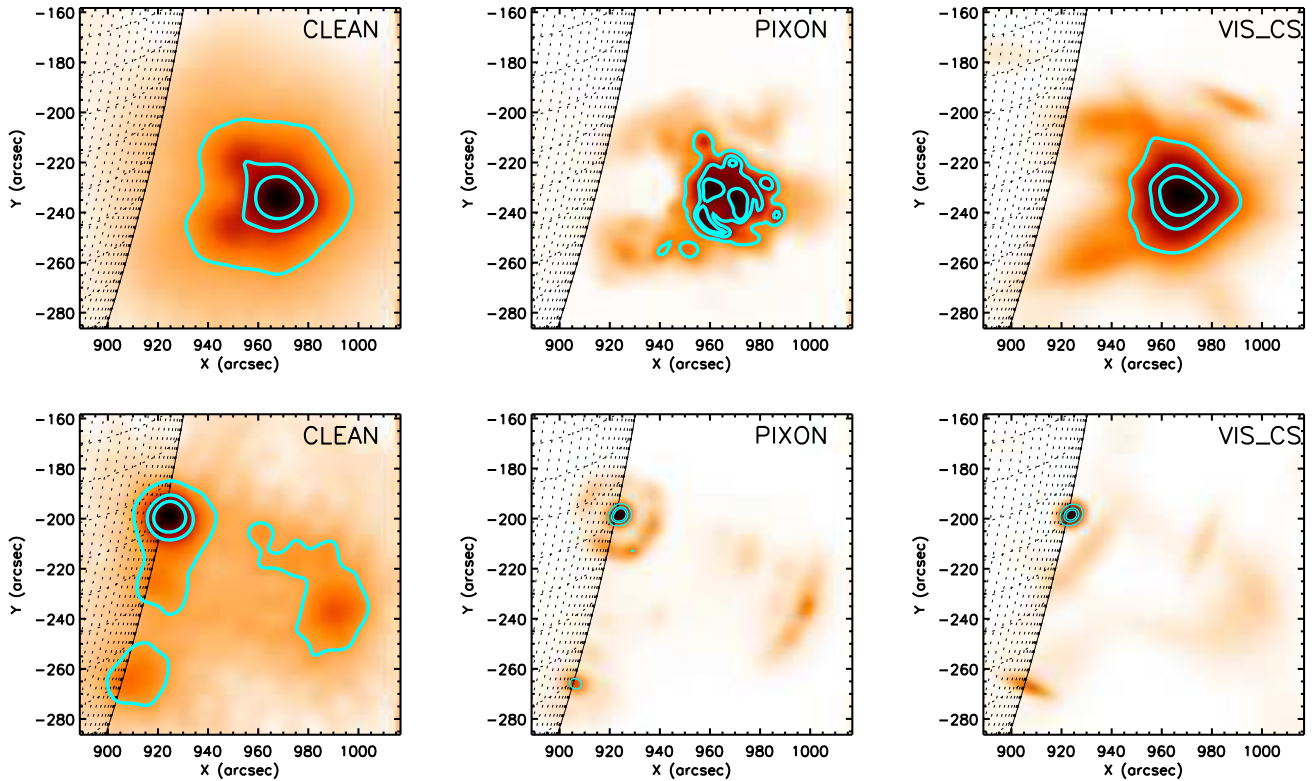}
\caption{Comparison of image reconstructions of two types of actual flare sources from three different algorithms. Left to right: Clean, Pixon, VIS\_CS. Top row: extended source above the solar limb. Bottom row: one bright and one faint compact source. The contours give the 20\%, 50\% and 70\% level relative to the maximum emission in each image. VIS\_CS reconstructs the extended source without fragmentation and is capable of recovering compact sources even without source-specific detector selection.}
\label{fig_ext_compact}
\end{figure*}

For our comparison, we chose an event that occurred on 19 July 2012 with a high energy X-ray peak-time around 05:22 UT. The event is described in detail by e.g. \citet{Bat13} and \citet{Kru}. During the peak intensity of the emission, it displayed an extended source above the solar surface and two compact sources close to the solar surface. We used two standard image reconstruction algorithms (Clean and Pixon). For these, detectors 3-8 were used, but without detector 4 as it was excessively noisy at the time. For all other parameters, the defaults were used. The VIS\_CS image was made using all detectors, except detector 4. The images were made for a one minute integration time from 05:21 UT to 05:22 UT. Results are shown in Figure~\ref{fig_ext_compact}.

The extended source (imaged over energy range 6-12 keV) is well reproduced by Clean and VIS\_CS, with the advantage that VIS\_CS does not have any side-lobes. The Pixon image is unphysically fragmented. \revrem{This}\revadd{Both effects} could be \revrem{improved}\revadd{reduced or probably even eliminated} by fine-tuning algorithm parameters, which requires \revrem{Pixon-specific}\revadd{algorithm-specific} expertise.

The compact sources (imaged over energy range 25-50 kev) are well reproduced by all three algorithms. Here, Clean results in larger source sizes, again something that could be improved by fine-tuning parameters. Pixon reconstructs very compact sources but shows artifacts around them. All reconstructions show a weak residual emission in the corona. This emission, however, is at the noise level.

\subsubsection{Loops}
\begin{figure*}
\includegraphics[width=0.98\textwidth,trim={0 60px 0 60px},clip]{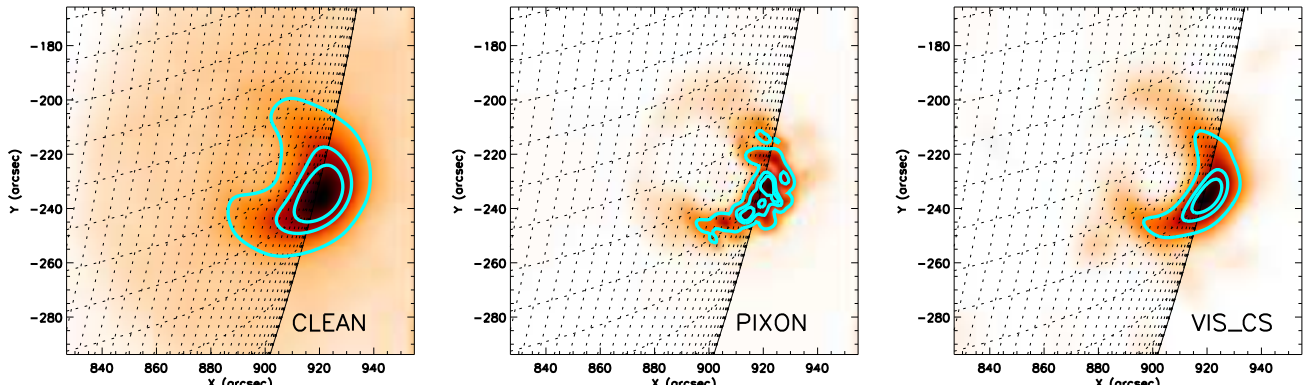}
\caption{Comparison of image reconstructions of a flaring loop from three different algorithms. Left to right: Clean, Pixon, VIS\_CS. The contours give the 20\%, 50\% and 70\% level relative to the maximum emission in each image. VIS\_CS recovers the loop-shape better than Clean without fragmenting the source.}
\label{fig_loop}
\end{figure*}

Although loops are very common in ultraviolet images of solar flares, \emph{RHESSI} images often only show the loop-top, most likely because imaging based on rotating modulating collimators is most sensitive to the brightest source in the field of view, making it difficult to observe co-temporal fainter sources. Some observations of X-ray loops were reported by
\citet{Nat13} who studied spatial and spectral properties of three X-ray loops by reconstructing the examined loops with various imaging algorithms. \revadd{Here we use one of the events analyzed by \citet{Nat13}. The event occurred on 23 August 2005 and we integrate over a two minute interval between 14:28:00 UT and 14:30:00 UT and over the energy range 10-12 keV}. Obtaining good reconstructions is difficult because of the low count rate. We perform the same reconstruction with VIS\_CS, using all detectors and compare the results with Clean and Pixon reconstructions on the same data (Figure~\ref{fig_loop}). For Clean and Pixon, detectors 3-8 were used with no additional imaging parameter optimization. The shape of the loop reconstructed with VIS\_CS is more clearly and concisely visible than the one reconstructed with Clean. 

\subsubsection{Above-the-looptop source}
\label{chap:above_looptop}
Flares typically show one coronal source observed at low energies (up to $\sim20$ keV), although signatures of accelerated electrons are expected at higher energies. Since these signatures tend to be rather faint compared to the bright low-energy coronal emission, they are difficult to observe with current instrumentation, as indicated above. The first observation of such an above-the-looptop source was made by \citet{Mas94}.

\begin{figure*}
\includegraphics[width=0.98\textwidth,trim={3px 90px 0 90px},clip]{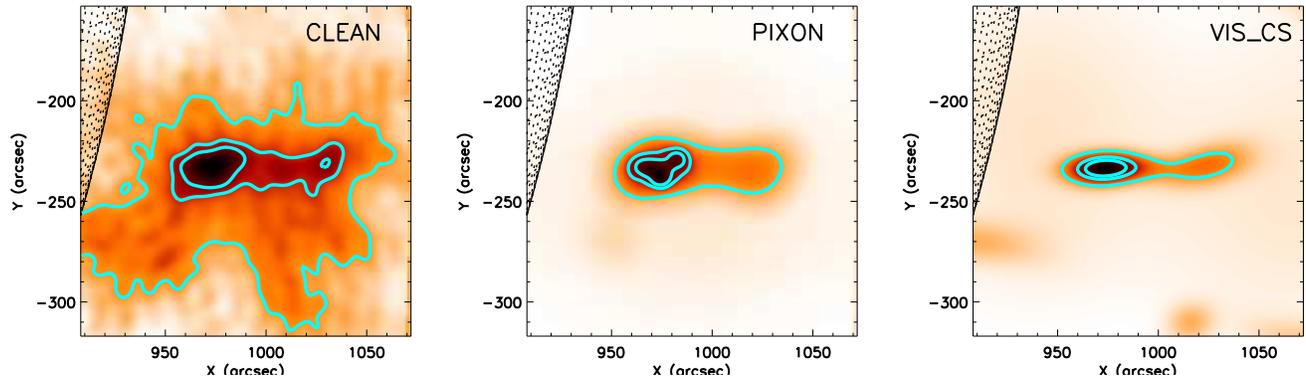}
\caption{Comparison of image reconstructions of two coronal sources from three different algorithms. Left to right: Clean, Pixon, VIS\_CS. The contours give the 20\%, 50\% and 70\% level relative to the maximum emission in each image. VIS\_CS is capable of reconstructing, and separating the faint above-the-looptop.}
\label{fig_looptop}
\end{figure*}

Before the event of 19 July 2012 described above, there was prolonged activity (up to 40 minutes before the actual flare) during which two separate sources could be identified in the corona \citep{Liu13,Sun14}. This presents an interesting science case allowing to study electron energization near the reconnection region, but requires a reliable imaging algorithm that is able to reconstruct the faint above-the-looptop source and separate it from the coronal source. We present the image reconstruction for a 3-minute time interval \revadd{from 04:46:00 and 04:49:00}, again comparing Clean with VIS\_CS (Figure~\ref{fig_looptop}), using detectors 3 and 5-8 over the energy range 16-20 keV for Clean and Pixon. For VIS\_CS, again all detectors were used except detector 4. While it \revrem{is}\revadd{would be} possible to \revadd{more clearly} separate the two sources by fine-tuning the Clean parameters, the \revrem{separation is more pronounced with VIS\_CS}\revadd{source separation in the default VIS\_CS reconstruction is more pronounced}. VIS\_CS also gives slightly elongated sources.

\section{Conclusions}
Image reconstruction of X-ray solar flare emission observed with instruments like \emph{RHESSI} requires specialized algorithms, each of which has advantages and disadvantages, depending on source morphology and application.
We systematically tested and demonstrated the ability of our new Compressed Sensing-based algorithm VIS\_CS to accurately reconstruct source locations, photometry, and morphology in a wide range of scenarios. The algorithm performs well in synthetic tests, as well as with observed flare data. The key advantage of VIS\_CS over the state-of-the-art is the ability to produce high-quality image reconstructions without algorithm- or source-specific parameter tuning and in a short time. Its robustness and speed mean that VIS\_CS is well-suited for applications where quick and reliable image generation for a variety of source morphologies with a minimum of user interaction is needed, such as for generation of quicklook images and large image cubes for imaging spectroscopy. While developed for analysis of \emph{RHESSI} data, the algorithm is not restricted to \emph{RHESSI} visibilities. Early tests indicate that VIS\_CS is a competitive image reconstruction algorithm for the upcoming STIX instrument \citep{Kru13}, which uses a similar imaging principle but with a different uv-coverage. 
We encourage the use of of VIS\_CS whenever fast, reliable reconstruction of a large number of images is needed and for less experienced users of \emph{RHESSI} data.
We made the IDL function \texttt{vis\_cs} available as part of the SolarSoft SSW \citep{freeland1998data} package.

\acknowledgments
We would like to thank Andr\'e Csillaghy, Richard A. Schwartz, L\'aszl\'o I. Etesi, S\"am Krucker, Gabriele Torre and Lucia Kleint for their help, insightful discussions, and suggestions. \revadd{We highly appreciate the constructive feedback from the referee.}

\bibliographystyle{apj}
\bibliography{main}

\end{document}